\PassOptionsToPackage{table,xcdraw}{xcolor}

\documentclass[acmsmall,screen]{acmart}

\usepackage[camera]{dtrt}

\usepackage{listings}
\usepackage{appendix}
\usepackage{makecell}

\usepackage{epsfig,endnotes}
\usepackage{multirow}

\usepackage{amsmath,amsfonts}
\usepackage{algorithmic}
\usepackage{graphicx}
\usepackage{textcomp}
\usepackage{pifont}
\definecolor{codeauburn}{rgb}{0.43, 0.21, 0.1}
\definecolor{codegreen}{rgb}{0,0.6,0}
\definecolor{codegray}{rgb}{0.5,0.5,0.5}
\definecolor{codepurple}{rgb}{0.58,0,0.82}
 
\lstdefinelanguage
   [x64]{Assembler}     % add a "x64" dialect of Assembler
   [x86masm]{Assembler} % based on the "x86masm" dialect
   {morekeywords={CDQE,CQO,CMPQ,CMPXCHG16B,JRCXZ,LODSQ,MOVSXD,movl,addl,subl, %
                  POPQ,PUSHQ,SCASQ,STOSQ,IRETQ,RDTSCP,SWAPGS,MOVABSQ,LEAQ, %
                  rax,rdx,rcx,rbx,rsi,rdi,rsp,rbp,retq,callq,movq, %
                  r8,r8d,r8w,r8b,r9,r9d,r9w,r9b,r10,r11,r12,r15,enclu,rdgsbase,wrgsbase,ud2,mfence,lfence,cmova,cmovg,cmovl}} % etc.
                  
\lstdefinestyle{mystylenobox}{
    backgroundcolor=\color{white},   
    commentstyle=\color{blue},
    keywordstyle=\color{codeauburn},
    numberstyle=\tiny\color{codegray},
    stringstyle=\color{codepurple},
    basicstyle=\footnotesize,
    breakatwhitespace=false,         
    breaklines=true,                 
    captionpos=b,                    
    keepspaces=true,                 
    numbers=left,                    
    numbersep=-1pt,
    xleftmargin=.25in,
    showspaces=false,                
    showstringspaces=false,
    showtabs=false, 
    tabsize=2
}

\makeatletter
\let\@ORGmakecaption\@makecaption
\long\def\@makecaption#1#2{\@ORGmakecaption{#1}{#2}\vskip\belowcaptionskip\relax}
\makeatother

\newcounter{packednmbr}

\usepackage{amsthm}

\usepackage{listingsutf8}
\definecolor{mulberry}{rgb}{0.772,0.29,0.549}

\usepackage{tcolorbox}

\newtheorem{insight}{Insight}

\def\Snospace~{\S{}}

\acmJournal{PACMSE}
\acmVolume{3}
\acmNumber{FSE}
\acmArticle{0}  % TODO: Replace with actual article number from ACM
\acmYear{2026}
\acmMonth{7}
\acmDOI{10.1145/3808159}

\begin{document}

%don't want date printed
\date{}

%make title bold and 14 pt font (Latex default is non-bold, 16 pt)
\title{Characterizing Trust Boundary Vulnerabilities in TEE Container Systems: An Empirical Study}

%%% Authors %%%
\author{Weijie Liu}
\authornote{Both authors contributed equally to this research.}
\orcid{0000-0002-3054-766X}
\affiliation{%
  \institution{Nankai University}
  \country{China}
}
\email{weijieliu@nankai.edu.cn}

\author{Hongbo Chen}
\authornotemark[1]
\orcid{0000-0001-9922-4351}
\affiliation{%
  \institution{Indiana University Bloomington}
  \country{USA}
}
\email{hc50@iu.edu}

\author{Shuo Huai}
\orcid{0009-0003-7103-5163}
\affiliation{%
  \institution{Nankai University}
  \country{China}
}
\email{shuohuai@mail.nankai.edu.cn}

\author{Zhen Xu}
\orcid{0009-0008-1127-4097}
\affiliation{%
  \institution{Nanyang Technological University}
  \country{Singapore}
}
\email{zhen.xu@ntu.edu.sg}

\author{Wenhao Wang}
\authornote{Corresponding authors.}
\orcid{0000-0001-7294-2724}
\affiliation{%
  \institution{Institute of Information Engineering, Chinese Academy of Sciences}
  \country{China}
}
\email{wangwenhao@iie.ac.cn}

\author{XiaoFeng Wang}
\authornotemark[2]
\orcid{0000-0002-0607-4946}
\affiliation{%
  \institution{Nanyang Technological University}
  \country{Singapore}
}
\email{xiaofeng.wang@ntu.edu.sg}

\author{Danfeng Zhang}
\orcid{0000-0003-1942-6872}
\affiliation{%
  \institution{Duke University}
  \country{USA}
}
\email{danfeng.zhang@duke.edu}

\author{Zhi Li}
\orcid{0000-0002-9510-1888}
\affiliation{%
  \institution{Huazhong University of Science and Technology}
  \country{China}
}
\email{lizhi16@hust.edu.cn}

\author{Haixu Tang}
\orcid{0000-0001-8963-8155}
\affiliation{%
  \institution{Indiana University Bloomington}
  \country{USA}
}
\email{hatang@iu.edu}

\author{Zheli Liu}
\authornotemark[2]
\orcid{0000-0002-2984-2661}
\affiliation{%
  \institution{Nankai University}
  \country{China}
}
\email{liuzheli@nankai.edu.cn}

\renewcommand{\shortauthors}{W. Liu, H. Chen, S. Huai, Z. Xu, W. Wang, X. Wang, D. Zhang, Z. Li, H. Tang, and Z. Liu}

%\LARGE

% \author{}

% for IEEE, ACM, and others, we put abstract above \maketitle
\begin{abstract}

Trusted Execution Environments (TEEs) have become a cornerstone of confidential computing, attracting significant attention from academia and industry. To support secure and scalable application deployment on confidential clouds, TEE containers (Tcons) have been introduced as middleware to shield applications from malicious operating systems and orchestration layers while preserving usability.
In this paper, we present the first comprehensive analysis of Tcons, focusing on three critical layers: OS interfaces, encrypted I/O, and orchestration mechanisms. To enable systematic evaluation, we design \textit{TBouncer}, an automated analyzer that precisely exercises and benchmarks Tcon isolation boundaries.
Our study uncovers fundamental flaws in existing Tcons, leading to exploitable vulnerabilities such as code execution, denial-of-service, and information leakage. In total, we identify six attack vectors, twelve new bugs, and three CVEs.
These findings provide new insights into the underestimated attack surface of Tcons and highlight key directions for building more secure and trustworthy container solutions.

\end{abstract}




\maketitle

% for USENIX, we put abstract below
% \input{tex/0-abstract}

% Camera-ready: page numbers managed by acmart

%tex

\section{Introduction}

% The demand for scalable and robust data protection in computing has spurred significant advancements in Trusted Execution Environment (TEE) technologies over the past decade, particularly in CPU-based isolated execution. Leading examples include Intel’s Software Guard Extensions (SGX)~\cite{sgx_official}, AMD’s Secure Encrypted Virtualization (SEV)~\cite{sev_official}, and ARM’s TrustZone (TZ)~\cite{trustzone_official}. These technologies have been widely adopted for secure critical tasks such as password management~\cite{apple_passkey}, network traffic protection~\cite{schwarz2020seng}, blockchain~\cite{cai2021toward}, and privacy-preserving data analysis~\cite{jin2024elephants}.
Trusted Execution Environments (TEEs) have advanced rapidly over the past decade to support scalable and robust data protection through hardware-assisted isolated execution. Representative examples include Intel SGX~\cite{sgx_official}, AMD SEV~\cite{sev_official}, and ARM TrustZone~\cite{trustzone_official}. These technologies are increasingly adopted in security-critical applications such as password management~\cite{apple_passkey}, secure networking~\cite{schwarz2020seng}, blockchain systems~\cite{cai2021toward}, and privacy-preserving data analysis~\cite{jin2024elephants}.

Despite their potential, TEEs remain difficult to adopt in practice, largely due to limited compatibility with unmodified applications. For example, SGX typically requires developers to adapt applications using specialized SDKs~\cite{asylo_official}, incurring substantial engineering effort. To bridge this gap, \textit{TEE middleware} emerges as a crucial software layer, serving as an intermediary, hosting applications, and bridging the gap between TEE hardware and general-purpose software.

\vspace{3pt}\noindent\textbf{Containerizing TEE Applications}.
Despite the advancements in TEE middleware, their adoption remains challenging~\cite{paju2023sok}. One of the key barriers to the widespread use of confidential computing has been the difficulty of application development and deployment. 
Some TEE runtimes address this challenge by enabling unmodified applications/containers to run directly within a TEE environment, \textit{bypassing the need for binary rebuilding}.
Examples include Gramine~\cite{tsai2017graphene} (formerly called Graphene-SGX), Occlum~\cite{occlum_asplos20}, and Mystikos~\cite{mystikos_repo}.

Our study focuses on this class of middleware as \textit{TEE container} (\textit{Tcon}). While Tcons significantly improve usability, they also increase the complexity of the TEE software stack.

\vspace{3pt}\noindent\textbf{Analyzing Tcon Isolation}.
The emergence of Tcons has introduced the convenience and transparency of running native code directly. With the industry’s growing demand for large-scale confidential service deployment~\cite{deml2024secureLLMs}, such as for large language models, containerization provides clear advantages, including streamlined scheduling and on-demand orchestration~\cite{zobaed2023confidential}.
Tcons simplify TEE usage but complicate the software stack, raising concerns about their impact on TEE security. Our research surveys state-of-the-art Tcons from academia and industry using public sources (e.g., papers, developer guides, source code). We analyze their claimed security features, interfaces, and isolation principles, highlighting trust boundaries and undocumented issues. 
To demystify isolated execution in popular Tcons, we developed security benchmarks, called \textit{TBouncer}, for the systematic evaluation of all the interfaces that need to be protected.

% \vspace{3pt}\noindent\textbf{Contributions}. Our contributions are summarized as follows.

% \vspace{2pt}\noindent$\bullet$\textit{New findings}.  
% We conduct a systematic study of TEE containers from both academia and industry, uncovering a broad attack surface across three layers: OS interfaces, encrypted I/O, and orchestration mechanisms. We further derive a taxonomy of six attack vectors that expose fundamental weaknesses in Tcon isolation. In total, we discover twelve previously unknown vulnerabilities, including memory corruption, rollback, and denial-of-service (DoS) issues, three of which have been assigned CVE IDs. All findings were responsibly disclosed.

% \vspace{2pt}\noindent$\bullet$\textit{New insights}.  
% Our analysis reveals that while Tcons improve the usability and scalability of TEEs, they also introduce underestimated risks across all three layers, highlighting the need for stronger boundary design, broader interface coverage, and more rigorous enforcement of freshness, atomicity, and confidentiality guarantees.

% In summary, \autoref{sec:background} presents a survey and trend analysis of modern TEE middleware. \autoref{sec:trust_boundaies} analyzes the trust boundaries of TEE containers. \autoref{sec:fuzzer} introduces the design of TBouncer. \autoref{sec:findings} reports  empirical findings. \autoref{sec:tradeoffs} discusses lessons and mitigations. \autoref{sec:discussion} examines applicability and limitations, and \autoref{sec:conclusion} concludes.

\vspace{3pt}\noindent\textbf{Contributions}. Our contributions are summarized as follows.

\vspace{2pt}\noindent$\bullet$\textit{ Systematic characterization of Tcon trust boundaries}.  
We conduct a comprehensive survey of TEE middlewares from both academia and industry (\autoref{sec:background}). We examine their isolation boundaries and security mechanisms across TEE-OS interfaces, encrypted I/O stack, and orchestration layers (\autoref{sec:trust_boundaies}).

\vspace{2pt}\noindent$\bullet$\textit{ Automated boundary-driven security evaluation}.  
We design and implement TBouncer, an automated bidirectional analyzer that systematically exercises Tcon boundaries. TBouncer enables repeatable and scalable security evaluation by correlating enclave-internal execution with host-observable behavior (\autoref{sec:fuzzer}). Our tool is publicly released~\cite{code_release}.

\vspace{2pt}\noindent$\bullet$\textit{ Empirical findings and security insights}.  
Using TBouncer, we derive a taxonomy of six attack vectors and uncover twelve previously unknown vulnerabilities, including memory corruption, rollback, and denial-of-service (DoS) issues, three of which have been assigned CVEs. We report extensive empirical findings (\autoref{sec:findings}), distill lessons and mitigation strategies (\autoref{sec:tradeoffs}), and discuss applicability and limitations in broader TEE environments (\autoref{sec:discussion}).
\section{TEE Middleware: Survey, Trends, and Taxonomy}
\label{sec:background}

In this section, we first introduce the background of commodity TEEs on the cloud in ~\autoref{sec:background:tee} and the original SDK-based programming model in ~\autoref{sec:survey:sdk}.
We then elaborate on the paradigm shift towards TEE runtime in ~\autoref{sec:survey:runtime}, breaking down the inherent limitations of the TEEs' original programming models and categorizing seminal solutions.
After that, we explain the combination of containerization and TEE in ~\autoref{sec:survey:containers}.
Related work is summarized in \autoref{sec:related} at the end of this section.

\subsection{Hardware Trusted Execution Environments}
\label{sec:background:tee}

\vspace{3pt}\noindent\textbf{Process-based TEE}.
Intel SGX provides application-level isolation by creating protected memory regions, called enclaves, for secure computation. However, SGX offers limited support for multi-process operations (e.g., \verb|fork|) and requires developers to redesign applications to fit the restricted enclave environment, increasing the complexity of porting large and resource-intensive systems.

\vspace{3pt}\noindent\textbf{VM-based TEE}. 
Emerging platforms such as SEV~\cite{sev_official}, CCA~\cite{arm_cca_whitepaper}, and TDX~\cite{intel_tdx_whitepaper} provide virtual machine-level TEEs. By introducing a full kernel layer, these platforms offer better compatibility and more flexible memory management, reducing reliance on specialized TEE runtimes. In confidential VM (CVM) deployments, the threat model assumes adversaries control all external inputs, including device I/O and interrupts.

\subsection{From SDK to TEE Middleware}
\label{sec:survey:sdk}

Intel’s SGX SDK and its driver were among the earliest tools supporting enclave development in TEE environments~\cite{sgxsdk}. They provide low-level APIs and require developers to partition applications into trusted and untrusted components. This process is both technically challenging and labor-intensive, demanding careful design to maintain efficiency, correctness, and security. Moreover, language support is often limited, further constraining adoption.
Subsequent frameworks, including the Open Enclave SDK~\cite{openenclave_official} and Rust SGX SDK~\cite{wan2020rustee}, partially mitigate these challenges but still require familiarity with platform-specific models and substantial code refactoring~\cite{shanker2020evaluation}. While improving compatibility, these SDKs continue to impose a steep learning curve.

\begin{figure}[]
	\centering
	\includegraphics[width=.88\textwidth]{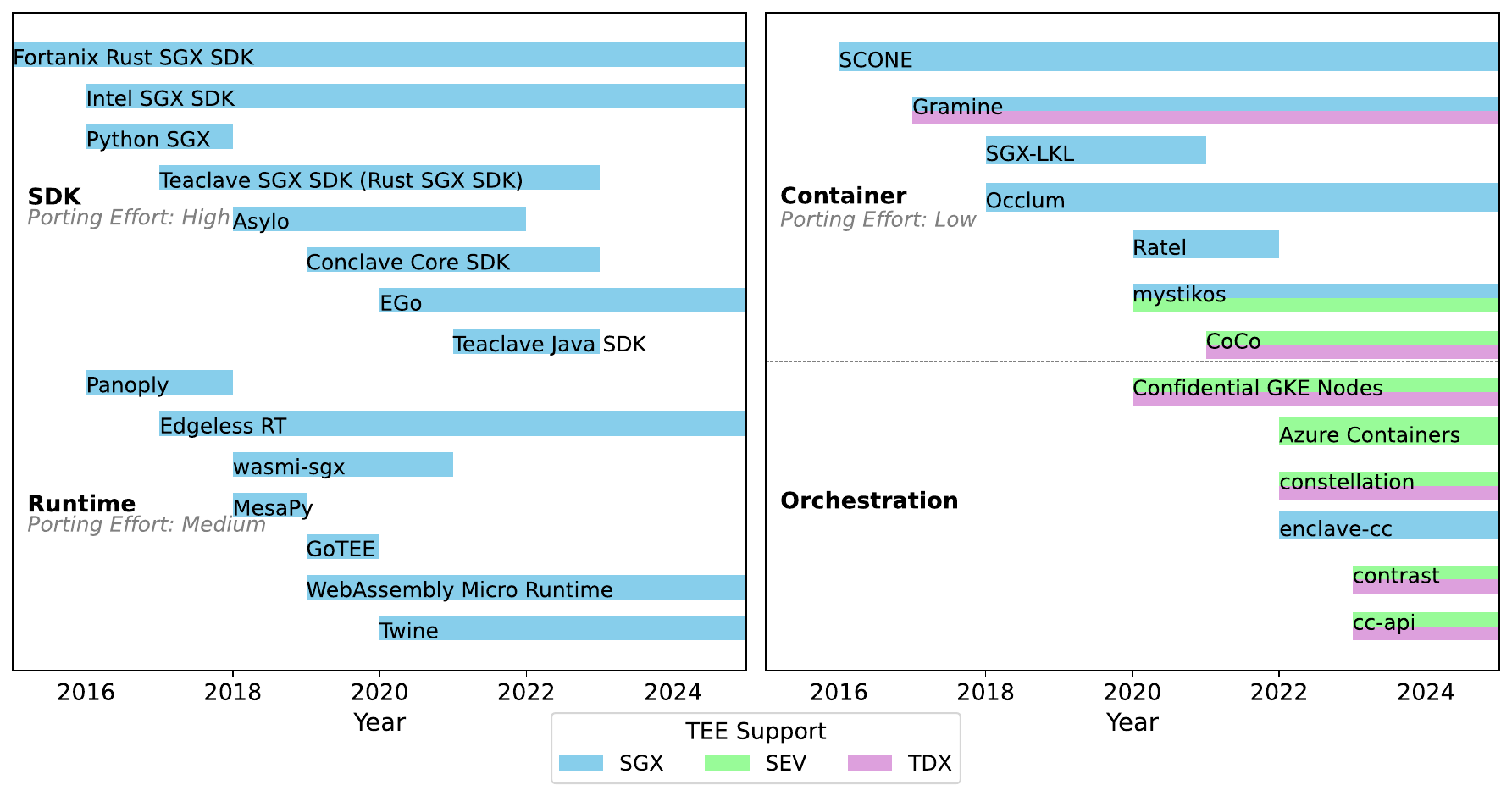}
	\caption{Timeline of TEE Middleware}
	\label{fig:timeline}
\end{figure}

\subsection{TEE Runtimes: Abstractions for Compatibility}
\label{sec:survey:runtime}

\vspace{3pt}\noindent\textbf{Software Compartmentalization}.
TEE architectures expose diverse host-OS interfaces, requiring SGX developers to handle these differences through often intrusive modifications. Solutions include compiler toolchains for rebuilding applications or runtimes that execute unmodified code.

\vspace{2pt}\noindent$\bullet$\textit{ Compiler toolchains}.
Frameworks such as Glamdring~\cite{lind2017glamdring} and Panoply~\cite{shinde2017panoply} automate trusted-untrusted separation using source-code annotations. Glamdring minimizes enclave transitions via static analysis, while Panoply isolates security-critical logic to reduce the TCB. However, both require manual annotations and substantial developer effort.

\vspace{2pt}\noindent$\bullet$\textit{ LibOS-based solutions}.
To ease integration, runtimes like Gramine~\cite{tsai2017graphene}, SCONE~\cite{arnautov2016scone}, Mystikos~\cite{mystikos_repo}, and Occlum~\cite{shen2020occlum} implement Library OSes (LibOS)~\cite{porter2011rethinking} that execute unmodified Linux binaries inside TEEs. These runtimes intercept system calls, forward them securely across enclave boundaries, and handle exceptions externally for compatibility and confinement. This design eliminates the need for manual partitioning, improving developer usability.

\vspace{3pt}\noindent\textbf{Limited Language Support}.
Early SDKs primarily targeted C/C++, offering limited support for other languages. Later systems introduced mechanisms for portability and memory-safe execution.

\vspace{2pt}\noindent$\bullet$\textit{ Language-specific runtimes}.
Frameworks like MesaPy and Teaclave SGX SDK~\cite{teaclave_sgx_sdk_repo} enable Python and Rust in SGX. MesaPy embeds a PyPy runtime with verified memory safety~\cite{wang2020towards}. These approaches often require enclave rebuilding, complicating deployment.

\vspace{2pt}\noindent$\bullet$\textit{ Wasm runtimes}.
Wasm sandboxes, e.g., AccTEE~\cite{goltzsche2019acctee}, Twine~\cite{menetrey2021twine}, and Enarx~\cite{enarx_official}, abstract enclave interactions via a language-neutral format. Wasm modules can be interpreted or compiled inside TEEs using engines like V8~\cite{v8_official} or WAMR~\cite{WAMR}, supporting SGX and SEV platforms.

\vspace{3pt}\noindent\textbf{Advanced Usage Scenarios}.
While early TEEs mainly targeted single-process workloads, modern deployments require multithreading, multi-user isolation, and secure concurrent execution.

\vspace{2pt}\noindent$\bullet$\textit{ Two-way sandboxes}.
Systems such as Chancel~\cite{ahmad2021chancel}, Ryoan~\cite{hunt2016ryoan}, and Occlum~\cite{occlum_asplos20} employ Software Fault Isolation (SFI)~\cite{tan_gang_sfi} to confine untrusted processes within enclaves. Occlum initially relied on Intel MPX, which was later removed due to deprecation~\cite{intelmanual}.

\vspace{2pt}\noindent$\bullet$\textit{ Executors and integration}.
Some runtimes act as modular executors within larger confidential computing frameworks. For example, Gramine and Occlum can be integrated into Teaclave, while Veracruz~\cite{veracruz_repo} and Oak~\cite{oak_repo} provide runtimes for Wasm-based workloads.

\subsection{TEE Containers}
\label{sec:survey:containers}

As confidential computing moves to the cloud, containerization has become essential. With recent platforms (e.g., SGXv2) supporting large enclave memory, TEEs can now host server-grade container workloads, motivating the exposure of TEE runtimes through container-compatible interfaces.

In this paper, we define \textit{TEE containers}, or \textit{Tcons}, as a category of TEE runtimes that are \textit{capable of accommodating applications without the need for code modification and recompilation}. According to the standard of the Open Container Initiative (OCI)~\cite{oci}, a Tcon should support deployment through a root filesystem (rootfs or image) and a configuration file.
As long as it complies with the OCI specification, it can be considered a container.
Projects like runc~\cite{runc} and Kata~\cite{katacontainer} integrate these mechanisms, establishing a widely adopted standard among OCI-compatible container runtimes.

\vspace{3pt}\noindent\textbf{Process-based integration}.
Gramine Shielded Containers, Occlum, and Mystikos containerize their LibOS runtimes, encapsulating enclave execution within standard container workflows. Each exposes filesystem and configuration metadata, enabling seamless deployment on Docker or Kubernetes. They've gained traction due to their minimal application intrusion and increasing maturity.

\vspace{3pt}\noindent\textbf{VM-based integration}.
The Confidential Containers (CoCo) project, including its Kubernetes operator~\cite{coco_operator}, integrates Kata Containers with VM-based TEEs (SEV/TDX). Each Pod runs inside a microVM, providing memory encryption and OS isolation~\cite{katacontainer}. While improving compatibility, this architecture enlarges the TCB and introduces challenges for secure storage and remote attestation.

\vspace{3pt}\noindent\textbf{Cloud-native extensions}.
Beyond standalone containers, recent efforts aim to integrate TEE containers into cloud platforms. Examples include Ahmad et al.’s orchestration framework~\cite{ahmad2023extensible}, Google Confidential GKE Nodes~\cite{confidential_gke}, and Azure Confidential Containers~\cite{azure_cc}, which support encrypted execution and confidential orchestration. Open-source projects such as Inclavare (enclave-cc)~\cite{inclavare_repo,enclave_cc} and Edgeless Contrast~\cite{constrast} further extend confidential computing to Kubernetes environments. Initiatives like CC-API~\cite{cc_api} seek to unify container interfaces across heterogeneous TEEs. Collectively, these efforts reflect the growing trend toward \textit{Confidential Kubernetes}~\cite{constellation}.

Figure~\ref{fig:timeline} summarizes the evolution of SDKs, runtimes, containers, and orchestration layers over the past decade. Our evaluation focuses on three \textbf{process-based Tcons} (Gramine, Mystikos, and Occlum) and one \textbf{VM-based Tcon} (CoCo~\cite{CoCo}), which support ``lift-and-shift'' deployment of unmodified applications.
Our survey covers active open-source projects up to Feb.~2026. Inactive, closed-source, or immature systems are excluded to ensure fairness and reproducibility. Table~\ref{tab:runtime-category} summarizes major TEE runtimes and containers.

\begin{table*}[]
\footnotesize
\caption{Summary of TEE Runtimes and Containers}
\label{tab:runtime-category}
\begin{center}
\begin{tabular}{|c|c|c|c|c|c|c|}
\hline
Category & Name & Mechanism & Platform & Activity & Usage & OS Interface \\ \hline

\multirow{2}{*}{\makecell{Compiler \\ toolchain}}
    & Glamdring~\cite{lind2017glamdring} & Annotation & SGX & \cellcolor[HTML]{C0C0C0}Inactive & \(\circ\)  & Generated stubs~\cite{lind2017glamdring} \\ \cline{2-7}
    & Panoply~\cite{shinde2017panoply} & Annotation & SGX & \cellcolor[HTML]{C0C0C0}Inactive & \(\circ\) & Panoply shim lib~\cite{shinde2017panoply} \\ \hline

\multirow{3}{*}{\makecell{Language \\ runtime}}
    & MesaPy~\cite{mesapy_repo} & Interpreter & SGX & \cellcolor[HTML]{C0C0C0}Inactive & \(\odot\)  & Intel-SGX-SDK~\cite{sgxsdk} \\ \cline{2-7}
    & AccTEE~\cite{goltzsche2019acctee} & Wasm & SGX & \cellcolor[HTML]{C0C0C0}Inactive & \(\circledcirc\) & SGX-LKL-OE~\cite{sgx_lkl_repo} \\ \cline{2-7}
    & Twine (WAMR)~\cite{menetrey2021twine} & Wasm & SGX & \cellcolor[HTML]{C0C0C0}Inactive & \(\circledcirc\) & Intel-SGX-SDK \\ \hline

\multirow{3}{*}{\makecell{Two-way \\ sandbox}}
    & Ryoan~\cite{hunt2016ryoan} & SFI & SGX & \cellcolor[HTML]{C0C0C0}Inactive & \(\circ\)   & NaCl loader~\cite{ryoan_code} \\ \cline{2-7}
    & Chancel~\cite{ahmad2021chancel} & SFI & SGX & \cellcolor[HTML]{C0C0C0}- (Closed) & \(\circ\)  & Intel-SGX-SDK \\ \cline{2-7}
    & Deflection~\cite{liu2021practical} & SFI & SGX & \cellcolor[HTML]{C0C0C0}Inactive & \(\circ\)  & Intel-SGX-SDK \\ \hline

\multirow{4}{*}{\makecell{Function \\ in FaaS}}
    & Teaclave executor~\cite{teaclave_function_executors} & Wasm & SGX/TZ & \cellcolor[HTML]{C0C0C0}Inactive & \(\circledcirc\)  & Teaclave-SDK~\cite{teaclave_sgx_sdk_repo} \\ \cline{2-7}
    & Enarx keeps~\cite{enarx_official} & Wasm & SGX/SEV &  Monthly & \(\circledcirc\)   & Enarx shim layer~\cite{enarx_official} \\ \cline{2-7}
    & Veracruz engine~\cite{veracruz_repo} & Wasm & SGX/TZ & Yearly & \(\circledcirc\)  & Teaclave-SDK \\ \cline{2-7}
    & Oak function~\cite{oak_repo} & VM & SEV/TDX & Weekly & \(\circledcirc\)   & VirtIO~\cite{virtio} \\ \hline

\multirow{8}{*}{Container} 
    & SCONE~\cite{arnautov2016scone} & SGX-aware libc & SGX & \cellcolor[HTML]{C0C0C0}- (Closed) & \(\bullet\)   & Ocall stubs~\cite{arnautov2016scone} \\ \cline{2-7}
    & SGX-LKL~\cite{sgx_lkl_repo} & LibOS & SGX & \cellcolor[HTML]{C0C0C0}Inactive & \(\bullet\)   & SGX-LKL-OE \\ \cline{2-7}
    & Ratel~\cite{cui2022dynamic} & DBT & SGX & \cellcolor[HTML]{C0C0C0}Inactive & \(\bullet\)   & Ratel-SGX-SDK~\cite{ratel} \\ \cline{2-7}
    & Gramine~\cite{tsai2017graphene} & LibOS & SGX & Weekly & \(\bullet\)   & Customized PAL~\cite{gramine_repo} \\ \cline{2-7}
    & Gramine-TDX~\cite{kuvaiskii2024gramine} & LibOS & SGX & Monthly & \(\bullet\)   & Specific VirtIO~\cite{gramine_tdx_repo} \\ \cline{2-7}
    & Occlum~\cite{occlum_asplos20} & LibOS & SGX & Monthly & \(\bullet\)   & Teaclave-SDK \\ \cline{2-7}
    & Mystikos~\cite{mystikos_repo} & LibOS & SGX & Monthly & \(\bullet\)   & OE SDK~\cite{openenclave_official} \\ \cline{2-7}
    & CoCo~\cite{CoCo} & VM & SEV/TDX & Weekly & \(\bullet\)   & VirtIO driver \\ \hline

\end{tabular}
\\
Usages: \(\circ\) Recompile into object/module, \(\odot\) Run specific language, \(\circledcirc\) Build and run Wasm, \(\bullet\) Run binary/image
\end{center}
\end{table*}

\subsection{Previous Works}
\label{sec:related}

Several studies analyze the usability and portability of TEEs. Shanker et al.~\cite{shanker2020evaluation} examine trade-offs in porting legacy applications to SGX, while Li et al.~\cite{mengyuan2024sok} and Paju et al.~\cite{paju2023sok} provide comprehensive surveys of TEE architectures, applications, and performance. Mo et al.~\cite{mo2024machine} review privacy-preserving machine learning in TEE environments. None of the above surveys systematically investigate TEE containers as a deployment paradigm.

Prior research has also explored boundary vulnerabilities in enclave-based systems. 
Van Bulck et al.~\cite{van2017sgx} discuss API/ABI level sanitization vulnerabilities in shielding runtimes on SGX, TrustZone, and RISC-V. They analyze the bridge functions with vulnerabilities that can lead to exploitable memory safety and side-channel issues.
Cui et al.~\cite{cui2021emilia} developed \textit{Emilia} to automatically detect Iago vulnerabilities in SGX applications by fuzzing applications using syscall return values.
Khandaker et al.~\cite{khandaker2020coin} introduced COIN attacks and proposed an extensible framework based on instruction emulation and concolic execution to assess enclave vulnerabilities. Cloosters et al.~\cite{cloosters2020teerex} developed \textit{TeeREX} to automatically apply symbolic execution to analyze enclave binary code.
Wang et al.~\cite{wang2023symgx} proposed \textit{SymGX}, which employs static symbolic execution to detect cross-boundary pointer vulnerabilities in SGX applications by tracking pointer propagation. 
Alder et al.~\cite{alder2024pandora} introduced \textit{Pandora} to apply principled symbolic validation to enclave runtimes, which uncover logic bugs and memory safety issues. 
These works primarily focus on SDK-based models and specific enclave interfaces. In contrast, our work targets the broader trust boundary exposed by TEE containers, including syscall interfaces and I/O paths, which arise from supporting unmodified applications.
\section{Trust Boundaries of TEE Containers}
\label{sec:trust_boundaies}

% Existing Tcon research lacks a systematic evaluation of trust boundary enforcement. We examine the isolation guarantees of representative Tcons, focusing on (i) externally exposed interfaces and (ii) their protection.
% Our study classifies Tcon boundaries into three primary types: OS interfaces, encrypted I/O stacks, and orchestration interfaces, each serving distinct architectural roles.
Existing Tcon research lacks a systematic evaluation of trust boundary enforcement. In this work, we examine the isolation guarantees of representative Tcons by focusing on (i) externally exposed interfaces and (ii) their protection mechanisms. Our study classifies Tcon trust boundaries into three primary types: OS interfaces, encrypted I/O stacks, and orchestration interfaces, each serving distinct architectural roles.
% As illustrated in Figure~\ref{fig:interface-design}, application interfaces include Application Binary Interfaces (ABIs), primarily Linux ABIs with some POSIX-compatible variants.
% OS interfaces enable runtime communication between the TEE and the host OS. In process-based Tcons, this involves Ocall-to-syscall chains. These interfaces handle critical tasks like secure I/O, but also pose risks, such as potential compromise of confidential I/O channels by a malicious host.

\begin{figure}[]
	\centering
	\setlength{\abovecaptionskip}{0.cm}
	\includegraphics[width=.9\textwidth]{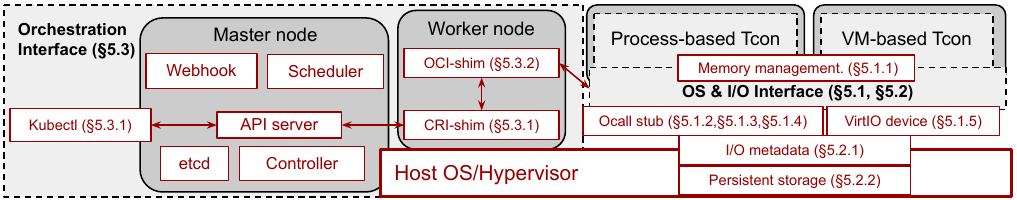}
	\caption{Interface of Tcons}
	\label{fig:interface-design}
\end{figure}

As illustrated in Figure~\ref{fig:interface-design}, application-level interfaces are mainly defined by Application Binary Interfaces (ABIs), which are primarily Linux ABIs with some POSIX-compatible variants. OS interfaces enable runtime communication between the TEE and the host OS. In process-based Tcons, this interaction is realized through OCall-to-syscall chains, while in VM-based Tcons it is mediated by paravirtualized devices such as VirtIO. These interfaces are essential for supporting secure I/O and resource management, but also constitute major attack surfaces that can be abused by a malicious host.
At the orchestration layer, Tcons are managed by container runtimes and Kubernetes components, which expose control and management interfaces for deploying, scaling, and monitoring confidential workloads. While these interfaces enable efficient resource utilization and flexible multi-tenant management, they also introduce additional trust dependencies that may undermine isolation if improperly protected.
Figure~\ref{fig:interface-design} provides a unified view of these interface layers and highlights the major trust boundaries examined in this paper. It further serves as a roadmap for the vulnerability analysis presented in~\autoref{sec:os_interfaces} - \autoref{sec:oi_interface}, where we systematically investigate security issues across these layers. Table~\ref{tab:boundary-interfaces} summarizes the corresponding interface categories.
% Orchestration interfaces facilitate container management in Kubernetes environments, exposing API endpoints for deploying, scaling, and managing TEE resources. These interfaces enable efficient resource utilization and secure workload management in multi-tenant setups.
% Table~\ref{tab:boundary-interfaces} summarizes those interfaces.

\begin{table*}[]
\footnotesize
\caption{Tcon Boundaries across OS interfaces, encrypted I/O mechanisms, and orchestration layers}
\label{tab:boundary-interfaces}
\begin{center}
\begin{tabular}{|c|c|c|c|c|c|}
\hline
\multirow{2}{*}{Tcon} & \multirow{2}{*}{OS Interface} & \multicolumn{2}{c|}{Encrypted I/O Stack} & \multicolumn{2}{c|}{Orchestration Control Interface} \\ \cline{3-6}
 &  & Disk I/O & Network I/O &  Runtime & Orchestrator \\ \hline

Gramine  & Gramine PAL & Gramine PF~\cite{gramine_pfs} & Rakis~\cite{alharthi2025rakis} & rune~\cite{rune} & enclave-cc~\cite{enclave_cc} \\ \hline

Occlum   & Ocall stubs & Async-SFS~\cite{occlum_async_sfs} & $*$ & rune & enclave-cc \\ \hline

Mystikos & Ocall stubs & \begin{tabular}[c]{@{}c@{}}ext2fs~\cite{mystikos_ext2fs} (input protection)\\ hostfs~\cite{mystikos_hostfs} (output persistence)\end{tabular} & $*$ & - & - \\ \hline

CoCo     & VirtIO driver & dm-crypt~\cite{dm_crypt} & $*$ & kata-shim~\cite{katacontainer} & coco-operator~\cite{coco_operator} \\ \hline

\end{tabular}
\\
$*$ No special handling is required.
\end{center}
\end{table*}

% gramine: https://carteepaul.github.io/gramine/devel/features.html?highlight=syscall#list-of-system-calls
% mystikos: https://github.com/deislabs/mystikos/blob/main/include/myst/syscall.h#L125
% how to containerize: https://github.com/deislabs/mystikos/issues/736

\subsection{OS Interfaces}

% The trust boundary between the host and the TEE can be placed at various levels, exposing different trade-offs that influence security, performance, and compatibility.

% \vspace{3pt}\noindent\textbf{ABIs}.
% Tcons mediate binary interactions with the OS through a wrapped LibOS or a VM, managing system calls and interactions with pseudo-filesystems like \textit{procfs} and \textit{sysfs}. A Tcon can handle certain syscalls internally (e.g., \texttt{getpid}) or replace them with equivalent alternatives, effectively ``squashing'' or ``distorting'' syscalls to ensure proper application execution.
% Since SGX prohibits privileged instructions, invoking one triggers a \#UD exception, leading to Asynchronous Enclave Exits (AEX). 

\vspace{3pt}\noindent\textbf{SGX-specific Interfaces}.
While various TEEs provide distinct isolation granularities depending on their architectural boundaries, interfaces like Ocall, Ecall, and exception handling are unique to SGX.
% These interfaces facilitate interactions between trusted and untrusted components within SGX enclaves, where isolation is enforced at the enclave level, separating secure code from untrusted host code.
User-land enclaves lack the capability to perform tasks like direct I/O or privileged memory mapping. To bridge the gap, process-based Tcons use Ocall stubs, which are wrapper functions enabling enclaves to invoke untrusted host services, and exception handlers, which manage events such as asynchronous exits.
% \hongbo{``external calls'' might be a bit confusing: not sure if the call is from external or to external. Besides, what exactly are ``ocall stubs''?}
This layer serves as a lightweight mediator, delegating privileged operations to external handler libraries.
% \hongbo{seems like something is missing here. What about exitless Ocall? Those are based on direct memory write and may not be interpreted as Ocall? Or we don't need to explain it?}
Note that exitless Ocall mechanism~\cite{yuhala2023sgx} is implemented in some Tcons for performance enhancement~\cite{occlum_ngo}. 

Tcons heavily rely on Ocall stubs to request kernel services, though the degree of reliance varies across different Tcons. %Mystikos and Occlum, for example, adopt distinct approaches to syscall mediation in secure execution environments.
For example, Mystikos uses \textit{Tcall}~\cite{tcall}, a mechanism that directly invokes Linux syscalls. In contrast, Occlum processes most syscalls internally within its LibOS, treating these exceptions as custom operations. Once processed, these operations are forwarded to the host kernel for execution. This reflects their differing philosophies: Mystikos emphasizes direct syscall execution complemented by external security measures, while Occlum prioritizes internal exception handling and syscall abstraction within the enclave.
Besides, Tcons differ in their support for raw syscall assembly instructions.
Gramine routes raw syscalls and exceptions through its platform abstraction layer to the kernel. Occlum processes these exceptions internally as custom syscalls before forwarding them to the kernel.

\subsection{Encrypted I/O Stacks}

% \vspace{3pt}\noindent\textbf{I/O Interfaces}.
Figure~\ref{fig:layers} shows the I/O stacks of Linux storage and network.
With networking, the interfaces can be sockets, TLS-encrypted TCP stack, etc. For storage, it can be file I/O and block I/O.
% but we actually need to discuss what layers should be boundaries

% IOfigure: https://arxiv.org/pdf/2403.03360

% for coco，its trust boundary is at the virtio layer。

% Virtio-net operates at the second layer of the OSI model, namely the Data Link Layer (Layer 2). Virtio-net is a paravirtualization technology designed to enhance the performance of network devices within virtual machines. It achieves efficient data transfer by establishing virtual queues (virtqueues) between the virtual machine (guest) and the host machine (host). These virtual queues are used for sending and receiving data packets, thereby avoiding the performance overhead associated with traditional virtualization technologies. The front-end driver of Virtio-net runs in the kernel space of the virtual machine, while the back-end driver runs in the kernel space of the host machine, with both communicating through the virtio layer and the virtio-ring layer. Therefore, Virtio-net primarily handles the functions of the Data Link Layer, including the transmission of Ethernet frames.

\vspace{3pt}\noindent\textbf{Disk I/O and Secure Storage}.
Confidential disk I/O is critical in the design of TEE containers, given the Unix-like principle that ``everything is a file''. These interfaces must be carefully integrated into the TEE framework to ensure compatibility and security. In SGX-based Tcons, Disk I/O operations are mediated by Ocalls. Conversely, in VM-based Tcons leveraging technologies like TDX and  SEV, Disk I/O often relies on VirtIO interfaces, such as virtio-blk, to facilitate interactions between the guest OS (Trust Domain/Secure VM) and the underlying hardware.

% Ensuring confidentiality and integrity of persistent and ephemeral storage is pivotal. Tcons leverage the sealing capabilities of their respective TEEs to safeguard sensitive data. 
% \hongbo{Some content overlaps sec 6, might be reduced if exceeds page limitation.}
Tcons adopt diverse approaches to secure storage. For instance, Occlum provides a protected file system to shield data at rest, while Gramine introduces APIs for transparent encryption and decryption of file operations, minimizing the burden on developers.
Mystikos currently supports three types of file systems: ramfs, ext2fs, and hostfs. 
% Ramfs and hostfs lack encryption and integrity checks, rendering them vulnerable. 
%
For CoCo, secure storage is mainly implemented by integrating dm-crypt~\cite{dm_crypt} through cryptsetup~\cite{cryptsetup} as the encryption mechanism. 
% By leveraging dm-crypt and dm-integrity~\cite{}, CoCo ensures both confidentiality and integrity of block volumes. 
These encrypted volumes, compatible with Kubernetes PodSpecs, are integrated into container workloads via a pause sidecar container. This sidecar handles initialization and mounts the encrypted block volumes, streamlining deployment without requiring modifications to the user’s Pod definitions. 
The encrypted storage is used for hosting container images, root filesystem writable layers, and other ephemeral data.
% Additionally, CoCo employs  (Linux Unified Key Setup) for block device encryption, providing a unified approach for both ephemeral~\cite{} and external persistent storage~\cite{}.

\vspace{3pt}\noindent\textbf{Network I/O}. 
% \hongbo{seems like this part is not connected with findings in Sec. 6.}
Tcons vary in their support for networks, with some relying on their own encryption mechanisms. 
% Attestation is crucial and often requires secure provisioning to transfer sensitive data to remote TEEs via encrypted network I/O\wenhao{check this sentence: attestation requires secure provisioning ...}.
Some Tcons only provide support at the I/O Multiplexing layer,
and some opt to encrypt network packets within applications rather than using confidential network I/O, avoiding the need to modify the I/O stack but incurring additional performance costs. The current design of such interfaces is suboptimal, with LibOSes and guest VMs increasing the TCB size and still requiring I/O to pass through the host, leading to potential vulnerabilities and performance bottlenecks~\cite{bifrost}. 
% \weijie{talk about SGX-LKL's design philosophy}

From a systems perspective, different methods for secure network I/O map to distinct layers of the OSI model~\cite{lefeuvre2023towards}. For instance, process-based Tcons' interfaces often operate at Layer 5 (the session layer), leveraging socket abstractions provided by the operating system. These interfaces are inherently tied to TCP/UDP traffic but depend on the host for low-level processing. 
On the other hand, VM-based Tcons transfer network packets via the virtio-net interface.

% High-performance solutions are also proposed~\cite{sev_tio,li2024bridge}, by contrast, bypassing these abstractions and working directly at Layer 2 (the data link layer), exchanging raw Ethernet frames through a dedicated TEE-managed I/O stack.

\subsection{Orchestration Control Interfaces}

The Kubernetes Control Plane orchestrates cluster resources and maintains the system’s desired state. Key components like the API Server, Controller, and Scheduler manage and optimize containerized application deployment. The Container Runtime Interface (CRI) abstracts the Kubelet from specific container runtimes, enabling Kubernetes to interact with diverse runtimes via gRPC-defined interfaces. This supports OCI-compliant runtimes like cri-o and cri-containerd.

Kubernetes Pods can be seamlessly replaced by CoCo or other Tcons (e.g., Gramine and Occlum) using solutions like CoCo's operator~\cite{coco_operator} or enclave-cc, a process-based runtime that manages TCon instances in Kubernetes.
Each Tcon has its own container-style runtime approach. Gramine implements Gramine Shielded Containers~\cite{gsc}, while Mystikos provides configurations through ``config.json''. Occlum adopts a distinctive approach by introducing an initfs mechanism to manage its root filesystem. Through the control plane, the CRI, and the OCI, tenants can use \verb|kubectl|~\cite{kubectl} to operate Tcons within their clusters. 
% For instance, Occlum’s deployment workflows on platforms like Azure Kubernetes Service leverage predefined interfaces for secure enclave initialization and runtime management~\cite{}.

% By enabling runtime flexibility and confidentiality at the orchestration layer, these interfaces ensure seamless integration with Kubernetes, while maintaining the security and isolation guarantees demanded by sensitive workloads.
% https://github.com/occlum/occlum/blob/master/docs/rune_quick_start.md
% https://github.com/occlum/occlum/blob/master/docs/azure_aks_deployment_guide.md

\subsection{Internal Isolation}

\vspace{3pt}\noindent\textbf{Sandboxing Malicious Code}.
Unmodified code might be vulnerable or even malicious, necessitating robust protection from Tcons.
This is particularly relevant in scenarios where service providers deploy proprietary code on cloud, and data owners, who are unable to audit the code, are concerned about potential privacy breaches~\cite{zhang2025erebor}. 
In response, certain TEE executors aim to safeguard data confidentiality by sandboxing unmodified code.
For example, the initial version of Occlum~\cite{occlum_asplos20} implemented SFI to secure hosted code, while Deflection~\cite{liu2021practical} introduced additional data protections over SFI. 
We found that the verifiers protect against indirect jumps but fail to adequately check direct jumps, which could be exploited to bypass security controls~\cite{code_release}. This oversight could allow a malicious application to take over the enclave.

\vspace{3pt}\noindent\textbf{Multi-user/thread Support}.
A single TEE may also process data from multiple users who may not trust one another, necessitating robust intra-TEE isolation for different tenants. Thread isolation is critical, as the service code deployed within a Tcon, while not malicious, may contain vulnerabilities. Without proper thread isolation, a compromised thread could interfere with others, potentially escalating into a broader security breach~\cite{ahmad2021chancel}.
Some techniques have been proposed to build a reusable environment while ensuring non-interference between user threads within an enclave or a confidential VM~\cite{hongbo2023security,park2025pave}.
Approaches to address this have relied on binary instrumentation through modified compilation toolchains~\cite{zhao2023security} or hardware modifications~\cite{park2024taco}.
% Current Tcons have not addressed these issues due to poor usability, which should be taken seriously.

\vspace{3pt}\noindent\textbf{Minimizing TCB}.
VM-based TEEs introduce unique challenges, notably expanding the TCB to include the entire kernel image and firmware components (such as UEFI/OVMF). 
This expansion has motivated research prototypes to replace the conventional Linux kernel with more lightweight and secure alternatives~\cite{asterinas}. Additionally, the trust chain in VM-based TEEs extends only to the initial OS image, hindering the verification of the application's integrity~\cite{narayanan2023remote}. To address this, some solutions attempt to enable the creation of middleware within VM-based TEEs, isolating them from the feature-rich TEE OS~\cite{zhou2024verismo,wang2025road}.

% Veil: A Protected Services Framework for Confidential Virtual Machines

% Previous research has explored two prototypes for SGX enclaves to address this challenge.
% \hongbo{I worry liveries is not considered TCon.} 
% Liveries proposed that data owners feed inputs sequentially, resetting the execution enclave to a benign initial state, thus avoiding the overhead of restarting the entire enclave.
% Chancel developed a multi-client SFI, offering each client a per-thread memory region and shared read-only area through a customized toolchain.

% \hongbo{However, these prototypes are no longer active, likely due to limited industry focus on this specific application and the fact that virtual machine-based TEEs already provide intra-TEE isolation, making them more accessible for service providers.}

% \vspace{3pt}\noindent\textbf{Insufficient Branch Checking}.
% Given the importance of thread isolation, enclave technologies are designed to isolate and protect applications from unauthorized access. They implement SFI to ensure that dangerous instructions are checked for safety before execution. However, a critical weakness has been identified in the verification of direct branch instructions within these technologies. While indirect jumps are protected, direct jumps are not adequately checked, potentially allowing malicious applications to bypass security measures and gain unauthorized control over the enclave~\cite{}.

\section{Fuzzing TEE Containers}
\label{sec:fuzzer}

In this section, we present \textit{TBouncer}, a suite of security benchmarks specifically designed to uncover intricate issues across the above-mentioned interfaces.
Our analysis targets x86-based TEEs, which dominate cloud and serverless deployments. ARM-based solutions, including TrustZone and CCA, are not the main focus of our empirical evaluation due to their distinct architectures and deployment maturity. 
Nevertheless, some vulnerability patterns are still applicable for mobile TEEs. We discuss the relevance and applicability in section~\autoref{sec:discussion}.

We evaluated Gramine (v1.5), Occlum (v0.30), Mystikos (v0.13.0), and CoCo (v0.11.0) following official documentation. SGX-based TCons ran on an Intel i7-1065G7 with 32 GB RAM; CoCo on AMD EPYC 7543 (SEV-SNP) and Intel Xeon 8568Y+ (TDX) with 488 GB RAM. Orchestration used enclave-cc for Gramine/Occlum and Kata shim plus CoCo-operator for CoCo.

\subsection{Design and Implementation}

\begin{figure}[]
	\centering
    \includegraphics[width=.88\textwidth]{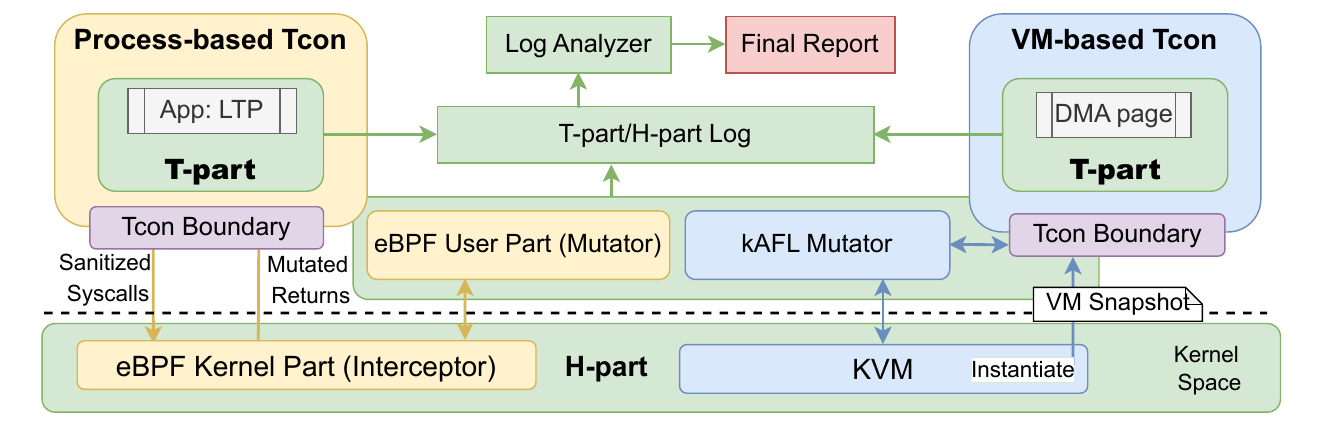}
	\caption{Workflow of TBouncer}
    % \hongbo{suggest denoting the operations in sequence.}
	\label{fig:fuzzer-pipeline}
\end{figure}

\vspace{3pt}\noindent\textbf{Two-part Framework}.
Our methodology centers around a two-piece analyzer comprising a T-part component within the Tcon and a host-level component (H-part) outside it.
This dual design enables test inputs to be injected either via the application running inside the Tcon or from the OS kernel, with responses monitored on the other end. 
Simply using tools like Trinity~\cite{trinity_repo} or AFL-derived fuzzers~\cite{kim2020hfl} can only test the Linux kernel unidirectionally, whereas our targets are fuzzing Tcons bidirectionally.
% https://blog.grimm-co.com/2020/05/analyzing-linux-kernel-in-userland-with.html
By simulating real-world interactions, our framework evaluates Tcon protections, such as input sanitization and return value handling. Both components operate independently to systematically identify vulnerabilities in Tcon runtime interactions.
The T-part generates diverse test cases targeting specific Tcon behaviors, while the H-part intercepts requests reaching the Tcon, inspects them, and can send back crafted responses to test the robustness of Tcons. 

As shown in Figure~\ref{fig:fuzzer-pipeline}, TBouncer uses eBPF~\cite{ebpf} for precise syscall interception and control, monitoring in-kernel syscall processing (e.g., parameters in \verb|sys_read| for \verb|read| calls)~\cite{jia2023programmable}. 
We then adapt kAFL~\cite{schumilo2017kafl} to simulate inputs sent to the bounce buffer from the host, enabling comprehensive Tcon-host interaction analysis. 

Despite the advantages of this 2-piece framework, one key challenge in analyzing Tcons lies in their strong isolation, while the host cannot perceive the application running within the TEE, complicating automated analysis. 
Kernel fuzzing tools such as Syzkaller~\cite{syzkaller_repo} also employ a two-part approach; however, they cannot be directly applied in this context due to two key limitations: (1) the inability to synchronize between the two components when the user-space part operates within a TEE, and (2) the lack of support for generating reverse test cases from the host to Tcons.

\vspace{3pt}\noindent\textbf{Precise Synchronization using Cgroups}.
To effectively monitor syscalls issued from T-part, filtering out interference from other host processes is essential. While one approach is to use eBPF to capture syscalls from specific process IDs (PIDs), this method has limitations, especially when multiple Tcon processes are running on the host concurrently. 
In practice, a Tcon may fork additional helper or child processes, meaning that tracing only the initial parent PID is insufficient to capture the full execution context.

To identify different Tcon processes, TBouncer leverages cgroups. The cgroup feature, which ensures that all child processes forked from a parent remain in the same cgroup, enables Tbouncer to comprehensively track all processes associated with Tcon and their activities.
Using cgroups, TBouncer groups all processes associated with a Tcon into a clean, isolated cgroup. This eliminates the need for the H-part to track multiple PIDs individually, as child processes automatically inherit their parent’s cgroup. The H-part then focuses exclusively on monitoring behaviors belonging to this cgroup, simplifying and streamlining syscall/DMA mutation.

To facilitate communication between the isolated Tcon and the host, TBouncer leverages a shared volume, a common debugging mechanism existing in all Tcons~\cite{Gramine_sv,occlum_async_sfs,mystikos_hostfs}.
This shared volume ensures precise signal exchange between the T-part and H-part with a designated \textit{signal file}.

% For example, before making a syscall, the T-part writes its name to a designated \textit{signal file} within the shared volume. The H-part monitors this signal file and activates eBPF to trace syscalls from the Tcon’s cgroup. 

% In the mutation phase, the T-part prefixes the syscall name and writes it to the signal file. The H-part intercepts the syscall and applies mutation logic as necessary. 
% Upon completion, the T-part signals, allowing the H-part to finalize the process and detach the mutation logic. If a syscall is successfully mutated, the H-part records its result in a dedicated file within the shared volume for further inspection. 
% This method ensures accurate filtering of host interference and efficient analysis of Tcon behavior.

\vspace{3pt}\noindent\textbf{Logging and Analyzing}.
TBouncer’s logging module facilitates in-depth analysis, enabling detailed analysis of parameters and behavior. Using these traces, the analyzer identifies which input parameters may have weaknesses and examines whether the sanitization mechanisms are implemented. To achieve this, TBouncer’s logging module compares parameters captured from both the T-part and H-part. This ensures comprehensiveness, exposing potential bugs in both Tcons' input handling and output sanitization.

\subsection{Input Generation Strategy}
\label{subsec:fuzzer}

\vspace{3pt}\noindent\textbf{Generating Syscalls}.
Syscall handling in Tcons varies: some syscalls are processed entirely within Tcons, bypassing the untrusted OS; others are unimplemented, conditionally handled, or forwarded to the host OS with modifications. For instance, a LibOS might translate a \verb|read| syscall into \verb|pread| by appending an offset. 
To enable the host OS to perceive the application's behavior running inside the T-part more accurately,
We modified the Linux Test Project (LTP)~\cite{ltp}, adapted to various Tcons by resolving dependencies and compatibility issues.
And the T-part traverses syscall parameters and nested data structures, recursively dereferencing pointers to capture comprehensive information.
% Additionally, it expands some cases by exploring parameters that align with their types but exceed normal ranges, such as treating enums as arbitrary integers or mismatching buffer lengths to trigger boundary checks.\wenhao{the threat model seems unclear; untrusted enclave? my major concern is why use abnormal ranges, or if we use, is it complete in generating mutations}

\vspace{3pt}\noindent\textbf{Crafting Iago-specific Tests}.
Iago is an attack in which a malicious kernel manipulates syscall return values to trick a protected process into acting against its own interests~\cite{checkoway2013iago}.
In VM-based TEEs, we extend this concept to malicious hypervisors, which could hijack the DMA interface and inject malicious input.
TBouncer also generates test cases to evaluate Tcon's resistance to Iago attacks. 
We consider both injecting \textit{syscall return values} from the host kernel and \textit{VirtIO parameters} from the device side.
We introduce the concept of nested structuring, which is strategically employed to populate syscall arguments and virtqueue descriptors. Furthermore, we simulate the interactions occurring at the boundaries of these structures by adapting and leveraging kAFL~\cite{schumilo2017kafl}.
To mutate the values from the LibOS/VM to the host OS at specific layers, we leverage eBPF to hook into critical functions (e.g., \verb|sys_epoll_wait| and \verb|sys_readlink|) associated with each data structure.

\vspace{3pt}\noindent\textbf{Stateless Fuzzing}.
To ensure stability and prevent interference with subsequent testing cycles, TBouncer performs a thorough cleanup at the end of each cycle. This includes unmapping memory regions, closing file descriptors, and properly joining threads. These steps ensure that each fuzzing iteration starts from a consistent and isolated state, avoiding unintended side effects and improving the reliability of the testing process.

\subsection{Coverage and Performance}

% \hongbo{Show our syscall ``coverage'' table here for different Tcons.}

% \begin{table}[]
% \footnotesize
% \centering
% \caption{Interface Coverage of TBouncer across Different Tcon Implementations}
% \label{tab:interface-coverage}
% \begin{tabular}{|c|c|c|c|c|c|}
% \hline
% \multirow{2}{*}{Category} 
% & \multirow{2}{*}{Tcon} 
% & \multicolumn{2}{c|}{Syscall Interfaces} 
% & \multicolumn{2}{c|}{VirtIO Interfaces} \\ \cline{3-6}
% & & Covered / Total & Coverage (\%) 
% & Covered / Total & Coverage (\%) \\ \hline
% \multirow{3}{*}{Process-based}
% & Gramine  & 169 / 169 & 100.0 & -- & -- \\ \cline{2-6}
% & Occlum   & 158 / 163 & 96.9  & -- & -- \\ \cline{2-6}
% & Mystikos & 179 / 203 & 88.2  & -- & -- \\ \hline
% VM-based
% & CoCo & 248 / 248* & 100.0 & 5 / 5 & 100.0 \\ \hline
% \end{tabular}
% \\
% *Forwarded system calls through the guest kernel.
% \end{table}

Instead of measuring traditional code coverage, we evaluate TBouncer in terms of \textit{interface coverage}, i.e., the extent to which exposed trust boundary interfaces are systematically exercised. A detailed discussion on the limitations of conventional coverage metrics is provided in Section~\ref{sec:discussion}.

Note that Tcons typically support only a subset of Linux system calls and may additionally introduce customized system interfaces. 
For syscall interfaces exposed to the untrusted OS, our security benchmark covers nearly all supported syscalls in process-based Tcons: 
169/169 for Gramine, 158/(158+5) for Occlum, and 179/(179+24) for Mystikos. 
Here, the additional syscalls correspond to Tcon-specific extensions implemented within the LibOS. 
As these interfaces are not part of the standard Linux ABI and lack standardized specifications, 
TBouncer currently does not provide dedicated fuzzing models for them and therefore does not systematically exercise these interfaces.

For VM-based Tcons, we exercised all 248/248 forwarded system calls in CoCo. 
In addition, the current implementation of CoCo supports five VirtIO devices, namely virtio-blk, virtio-net, virtio-console, virtio-9p, and virtio-vsock. 
We tested all exposed device interfaces and achieved full coverage (5/5). 
Table~\ref{tab:interface-coverage} summarizes the results.

\begin{table}[t]
\footnotesize
\centering
\caption{Interface Coverage Achieved by TBouncer}
\label{tab:interface-coverage}
\begin{minipage}{0.58\linewidth}
\centering
\begin{tabular}{|c|c|c|c|c|}
\hline
Tcon 
& Gramine 
& Occlum 
& Mystikos 
& CoCo \\ \hline
Syscall (Covered/Total)
& 169/169 
& 158/163 
& 179/203 
& 248/248* \\ \hline
Coverage (\%)
& 100.0 
& 96.9 
& 88.2 
& 100.0 \\ \hline
\end{tabular}
\end{minipage}
\hfill
\begin{minipage}{0.38\linewidth}
\centering
\begin{tabular}{|c|c|}
\hline
Tcon 
& CoCo \\ \hline

VirtIO (Covered/Total)
& 5/5 \\ \hline

Coverage (\%)
& 100.0 \\ \hline
\end{tabular}
\end{minipage}
\\
{\footnotesize * Forwarded via guest kernel.}
\end{table}

Furthermore, we evaluate the practical performance in terms of fuzzing throughput and testing effort. 
All experiments were conducted on our Intel Xeon 8568Y+ workstation running a Linux 6.8.0 kernel as the host OS, with a fixed time budget of 2 hours for process-based Tcons and 24 hours for VM-based Tcons. This budget proved sufficient to comprehensively explore exposed trust boundary interfaces, as both fuzzing throughput and vulnerability discovery stabilized substantially earlier—typically within 1 hour for process-based targets and 12 hours for VM-based targets.

During fuzzing, TBouncer continuously exercised Tcon boundary interfaces and monitored executions for abnormal behaviors, such as crashes, assertion failures, and inconsistent outputs. Detected anomalies were automatically logged with execution traces.
In particular, for DMA-related failures, TBouncer ensured that relevant buffers were properly initialized before issuing the triggering commands. Vulnerabilities identified are reported in detail in Section~\ref{sec:findings}.
\section{Findings}
\label{sec:findings}

We analyzed the recorded execution traces to validate reproducibility and confirm genuine security vulnerabilities, which were responsibly disclosed to the corresponding Tcon developer communities for verification and remediation.
We next present our findings from three perspectives: Iago-style attacks on OS interfaces, insufficient I/O encryption, and issues in orchestration layers.

\subsection{Iago-style Issues and Beyond}
\label{sec:os_interfaces}

\begin{figure}[]
	\centering
	\setlength{\abovecaptionskip}{0.cm}
	\includegraphics[width=.88\textwidth]{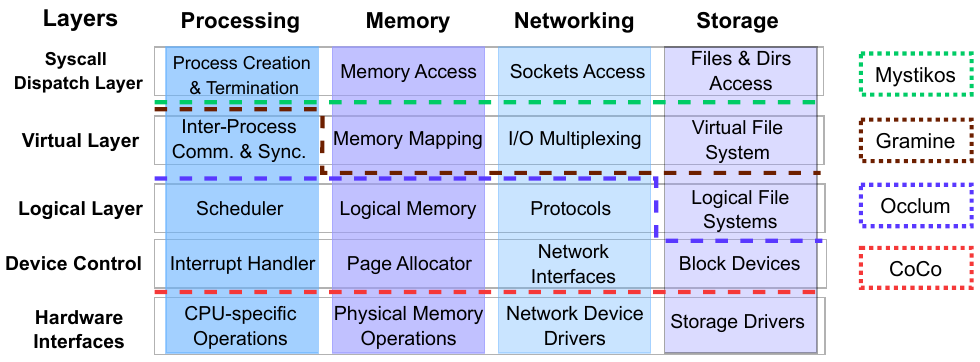}
	\caption{Layers of Tcon Implemented}
	\label{fig:layers}
\end{figure}

As analyzed in Section~\ref{sec:trust_boundaies}, issues arise when Tcons interact with the host OS, potentially exposing sensitive data or undermining the security model. 
We report vulnerabilities in Tcon's memory management subsystem, file I/O, inter-process communication (IPC), I/O multiplexing, and virtual drivers.
Figure~\ref{fig:layers} shows the existing protections in each part and highlights the boundaries and attack surfaces correspondingly.
To ensure the robustness of our findings, we engaged with the authors of related papers and developers of the selected Tcons. Their feedback helped validate our discoveries, refine interpretations, and identify key insights for the future evolution of Tcon technologies.
Table~\ref{tab:vulnerabilities} summarizes the issues.

% \usepackage{multirow}
% \begin{table}
% \footnotesize
% \caption{Summary of Vulnerabilities/Bugs in OS Interfaces}
% \label{tab:vulnerabilities}
% \centering
% \begin{tabular}{|c|c|c|c|c|} 
% \hline
% Tcon~~                & Memory Mgmt.   & File I/O \&  FS            & ~ ~IPC                & ~ ~I/O Multiplexing                  \\ 
% \hline
% Gramine~              & $\triangle$ ~           & Snooping~\cite{ahmad2018obliviate}          & ~ ~Insecure sems~\cite{code_release} (Bug-x) & $\triangle$ ~                                 \\ 
% \hline
% Occlum~~              & Iago~\cite{code_release} (Bug-1,2,3)     & ~ Improper return~\cite{code_release} (Bug-x) & DoS~\cite{code_release} (Bug-x)          & DoS~\cite{code_release} (Bug-x)                 \\ 
% \hline
% Mystikos~             & OOM crash~\cite{code_release} (Bug-x) & $\triangle$                      & $\triangle$                    & $\triangle$                                   \\ 
% \hline
% \multirow{2}{*}{CoCo} & \multicolumn{4}{c|}{VirtIO Driver}                                                                      \\ 
% \cline{2-5}
%                       & \multicolumn{4}{c|}{OOB access~\cite{code_release} (CVE-Vuln3)}  \\
% \hline
% \end{tabular}
% $\triangle$ = Partially supported by Tcons.
% $\bigstar$ = Iago (improper return or OOB access).
% $\blacksquare$ = DoS (Resource abuse or OOM crash).
% \\
% \end{table}

\begin{table}[]
\footnotesize
\caption{Summary of Vulnerabilities/Bugs in OS Interfaces.}
\label{tab:vulnerabilities}
\begin{center}
\begin{tabular}{|c|c|c|c|c|c|}
\hline
Tcon  & Memory Mgmt. & File I/O \& FS  & IPC & I/O Multiplexing & VirtIO Driver \\ \hline

Gramine &  $\triangle$  &  $\bigcirc$ Leakage \cite{ahmad2018obliviate}  & $\bigstar$ Iago (Bug-8) & $\triangle$  &  -   \\ \hline

Occlum       & $\bigstar$ Iago (Bug-1,2,3)  &  $\bigstar$ Iago (Bug-5,6) & $\bigstar$ Iago (Bug-9,10) & $\blacksquare$ DoS (Bug-11) & - \\ \hline

Mystikos        & $\blacksquare$ DoS (Bug-4)    &  $\triangle$  &   $\triangle$   &  $\triangle$  &  -    \\ \hline

CoCo        &  $\bigcirc$ Leakage \cite{cross2024ccs}   &   $\checkmark$   &  $\checkmark$ &  $\checkmark$  & $\bigstar$ Iago (CVE-1)\\ \hline
\end{tabular}
\\
$\bigcirc$: Information leakage: syscall snooping/memory disclosure.
$\bigstar$: Iago: improper return/insecure sems/OOB access.
\\
$\blacksquare$: DoS: resource abuse/OOM crash.
$\triangle$: Partially supported by Tcons.
$\checkmark$: Tested, but no abnormality observed.
\\
All Bugs and CVEs found by us are documented on our website~\cite{code_release}. Bug-x and CVE-x refer to vulnerability entries.
\end{center}
\end{table}

% or unintended file growth\hongbo{what's file growth?}.
% Another example is the Occlum's \verb|mprotect|. We found that the \verb|mprotect| incorrectly succeeds, allowing write permissions to be added to a read-only mapping~\cite{code_release}. In this way, sensitive files (e.g., configuration files, binaries, or logs) that are expected to remain read-only may be altered maliciously.
% \hongbo{Is is possible to extend these problems to some security implications? The following sections have the same problem.}
% \wenhao{this could lead to xxx. is it just a bug or is it a compromise that has to be made}

\subsubsection{Incomplete Memory Management}

Linux memory management is one of the kernel’s most intricate subsystems, supporting diverse functionalities. In SGX-based LibOS implementations, only a minimal subset is reimplemented to simplify design and maintain compatibility with unmodified applications. All Tcons enforce basic address space isolation (e.g., ensuring that returned addresses fall into enclave memory regions), but advanced support remains incomplete.

\begin{tcolorbox}[colback=white, colframe=black, boxrule=0.5pt, sharp corners]
\small
\textbf{Attack Vector 1: Syscall-level Iago Attacks.}
Tcons frequently exhibit semantic deviations or incorrect return values in system calls, enabling Iago-style attacks. Such deviations allow malicious hosts to subvert enclave semantics, resulting in corrupted data, and unauthorized access.
\end{tcolorbox}

Gramine implements core syscalls such as \verb|mmap|, \verb|munmap|, \verb|msync|, and \verb|mprotect|, but omits advanced flags and calls. For instance, \verb|MS_INVALIDATE| in \verb|msync|, and calls like \verb|mincore|, \verb|mlock|, and \verb|munlock| are unimplemented, which can cause performance degradation for applications relying on these features. Mystikos adopts an even stricter approach by enforcing configurable limits on stack and heap memory. Applications must be tuned to remain within these bounds, as exceeding them can result in crashes or out-of-memory errors~\cite{code_release}, underscoring the fragility of memory management in SGX-based environments~\cite{mckeen2016intel}.

Occlum, implemented in Rust, benefits from memory safety guarantees that reduce many low-level risks. However, our benchmark revealed vulnerabilities, illustrating how incomplete memory management can lead to Iago-style issues. For example, when \verb|mmap| maps a file, the memory area beyond the file’s end should be zero-filled. Instead, Occlum returned anomalous values~\cite{code_release}, allowing unintended modification of data beyond the file boundary, potentially corrupting file content. Similarly, \verb|mprotect| incorrectly succeeds in granting write access to read-only mappings~\cite{code_release}. This flaw enables malicious modification of sensitive files (e.g., configurations, binaries, or logs) that should remain immutable. 

In VM-based Tcons, adversaries can further exploit CoCo’s insufficient memory isolation to induce a wide range of attack effects. Prior work reports a cross-container memory disclosure attack in which a malicious container is deployed alongside a victim container~\cite{cross2024ccs}. Once the attacker container is launched within the same sandbox, embedded scripts enumerate the memory mappings of processes inside the victim container, dump their contents, and transfer them to the host file system for offline analysis. 
% Consequently, sensitive information such as application code, credentials, and encryption keys can be extracted from neighboring containers without violating coarse-grained isolation mechanisms.
This attack is primarily enabled by incomplete isolation of process visibility and memory management metadata inside the confidential VM, as well as overly permissive control interfaces that lack fine-grained authorization between co-located containers.

\subsubsection{Incomplete File I/O}

Gramine, Occlum, and Mystikos provide basic filesystem capabilities, supporting syscalls such as \verb|read|, \verb|write|, \verb|lseek|, \verb|open|, and \verb|close|. However, many advanced features remain unsupported or only partially implemented. For instance, event monitoring via \verb|inotify| is not available, while file linking and locking mechanisms (e.g., \verb|fcntl|) are restricted. Moreover, Gramine and Mystikos isolate each process in a separate enclave, thereby fragmenting filesystem metadata across enclaves. This design prevents dynamic filesystem mounting, makes cross-process file synchronization infeasible, and leads to non-interruptible file locking.

\vspace{3pt}\noindent\textbf{Semantic Deviations in Syscalls}. 
The \verb|readlinkat| syscall is expected to return an \verb|EINVAL| error when invoked with \verb|bufsiz| set to `0’. However, our security benchmark found that Occlum unexpectedly returned \verb|Success| under these conditions~\cite{code_release}. 
This anomaly introduces a risk of writing to an unallocated or improperly sized buffer, potentially leading to memory corruption.
Furthermore, this vulnerability could be exploited by attackers to perform DoS through repeated calls.
Similarly, the \verb|sendfile| is expected to return an \verb|EBADF| error if the output file descriptor lacks write permissions~\cite{code_release}. However, Occlum was observed to incorrectly return \verb|Success| in such cases.
This flaw could allow unauthorized writes to restricted files, bypassing access controls. These scenarios could potentially enable misuse by applications.

\subsubsection{Untrusted IPC}

Linux Inter-Process Communication (IPC) relies on mechanisms like memory sharing, semaphores, and message queues. Memory sharing provides access to the same physical memory across processes.
However, we found that the single-enclave-per-process architectures of Gramine and Mystikos introduce challenges for fully implementing IPC. 
Implementation of POSIX semaphores is insecure, as semaphores are placed in shared memory which by design is allocated in untrusted non-enclave memory, and there is no way for Gramine to intercept memory accesses to shared memory regions (to provide some security guarantees)~\cite{gramine_semaphores}.
Meanwhile, POSIX-based implementations such as GPU communication, remain unsupported in both systems.

\begin{tcolorbox}[colback=white, colframe=black, boxrule=0.5pt, sharp corners]
\small
\textbf{Attack Vector 2: Untrusted Shared Memory.}
Reliance on host-allocated shared memory exposes Tcons to tampering, leakage, and threatening system stability through manipulated IPC behavior.
\end{tcolorbox}

Achieving both robust isolation and high IPC performance in enclaves is challenging.
Gramine-SGX assigns an enclave to each process. Meanwhile, Gramine-TDX extends Gramine LibOS by incorporating a layer of \verb|vsock| and \verb|virtio_fs|~\cite{virtio_fs}, and it boots in the TDX environment. Consequently, when Gramine performs IPC, it needs to facilitate communication across enclaves or TDs, which incurs significant overhead.
% \hongbo{How bad is the performance? Why secruity affects the performance?}.
Occlum, with its single-enclave process model, consolidates data and different threads within the same enclave, allowing memory sharing. This design prioritizes performance.
Occlum's throughput on pipes is approximately 10 times that of Gramine~\cite{occlum_vs_gramine}.

Nevertheless, our tests still reveal several IPC limitations of Occlum. For example, \verb|shmget| inconsistently allocates shared memory sizes, while \verb|shmat| restricts the \verb|shmaddr| parameter to `0' or the starting address of a memory segment, disallowing manual address specification for security reasons. Additionally, the memory size reported by \verb|shmctl| may differ from the allocated size~\cite{code_release}. 
Such anomalies not only hinder compatibility with legacy applications but also introduce risks of undefined behavior, potentially leading to memory corruption or DoS in certain scenarios.
% Despite these shortcomings, Occlum includes return value checks in \verb|recvmsg|, a feature not implemented in some alternatives like CoCo.

\subsubsection{I/O Multiplexing and Event Handling}

Efficient management of concurrent I/O events is essential for high-performance and scalable systems. Linux provides three primary APIs for this purpose: \verb|select|, \verb|poll|, and \verb|epoll|. Both \verb|select| and \verb|poll| rely on linear data structures to track file descriptors, resulting in scalability limitations as the number of monitored sockets increases. In contrast, \verb|epoll| adopts an event-driven mechanism that notifies applications only when registered events are triggered, offering better performance for large-scale workloads.

\begin{tcolorbox}[colback=white, colframe=black, boxrule=0.5pt, sharp corners]
\small
\textbf{Attack Vector 3: Resource Exhaustion.}
Weak enforcement of resource limits and epoll misuse allow attackers to trigger deadlocks, exhaust resources, and induce DoS through crafted I/O patterns.
\end{tcolorbox}

Gramine and Occlum support all three I/O multiplexing APIs by emulating the host’s \verb|ppoll| syscall, allowing the implementation of \verb|select|, \verb|pselect|, \verb|poll|, and \verb|epoll| related syscalls. However, we identified several issues in their implementations.
In Gramine, the \verb|epoll| syscall does not handle the \verb|EPOLLWAKEUP| flag due to a lack of automatic resumption support. Additionally, Gramine restricts epoll usage: processes cannot share epoll instances, and nesting of epoll instances is unsupported due to enclave isolation. These limitations arise from Gramine’s focus on process isolation and security, sacrificing flexibility for enhanced security.
Occlum, on the other hand, has issues with epoll as well. It allows two epoll instances to listen to each other, causing deadlocks due to nested loops. 
Furthermore, there is no limit on epoll nesting, leading to potential resource exhaustion~\cite{code_release}.

These limitations involve both security and performance trade-offs. In Gramine, restricting epoll sharing and nesting limits inter-process communication, hindering scalability for complex I/O workflows. In Occlum, the lack of restrictions on epoll nesting introduces risks of deadlock and resource depletion. The design choices to limit or disable epoll sharing and nesting are intended to preserve enclave isolation but come at the cost of reduced performance and flexibility for high-concurrency, I/O-bound applications.

\subsubsection{Vulnerable Drivers and Device Interfaces}

VM-based TEEs are vulnerable to several security risks, particularly those arising from the exposure of device-shared pointers, Direct Memory Access (DMA) buffer control, and memory management flaws. These vulnerabilities, commonly associated with VirtIO drivers, present significant threats to system integrity and security. The following cases illustrate some of the specific risks and the Linux kernel patches designed to mitigate them.

\begin{tcolorbox}[colback=white, colframe=black, boxrule=0.5pt, sharp corners]
\small
\textbf{Attack Vector 4: Malicious DMA Manipulation.}
Paravirtualized drivers (e.g., VirtIO) expose enclave/VM pointers and rely on DMA descriptors from the untrusted host, causing out-of-bounds writes or code execution. This attack can be viewed as \textbf{a VM-level variant of Iago attacks}.
\end{tcolorbox}

\vspace{3pt}\noindent\textbf{Uncontrolled Device-accessible Pointers}.
In the network interfaces, a driver's pointer could be shared with an untrusted device. However under the threat model of TEE, when exposed to the untrusted hypervisor, it reveals the virtual machine’s randomized address space. This exposure opens the door for malicious overwrites from the host OS/hypervisor, as an attacker could potentially compromise the pointer and redirect execution by crafting \verb|sk_buff| objects, which can be fuzzed out by TBouncer. Specifically, when performing DMA memory access within Tcon, VM snapshots are taken.
TBouncer pinpointed the page and continuously randomized the data returned by the host.
By exploiting kernel function pointers, an attacker could then gain unauthorized code execution and then commit a practical attack~\cite{via2021acsac}.

% \ignore{
% The \verb|virtio_net| driver, in particular, can fail to handle initialization errors correctly, resulting in dangling pointers in freed memory. A malicious hypervisor could exploit this by forcing an invalid Maximum Transmission Unit (MTU) value, triggering memory reuse vulnerabilities. This flaw enables attackers to overlap freed \verb|virtnet_info| structures, manipulate function pointers, and escalate privileges. Recent patches address this issue by ensuring proper error handling and validating MTU values to prevent unintended pointer reuse~\cite{}.
% }

\vspace{3pt}\noindent\textbf{Out-of-Bound Access}.
Another common vulnerability stems from DMA buffer overflows. When device-managed DMA descriptors are not properly validated, a malicious device can inject arbitrary values into the descriptor ring, potentially causing buffer overflows. This exploitation can adversely affect the VirtIO driver. 
To find such issues, TBouncer tracks in-buffer lengths and injects data that exceeds these limits.
For example, we found that a lack of validation for the \verb|hash_key_length| parameter in the \verb|virtnet_probe| function can lead to out-of-bounds errors~\cite{code_release}. Various critical metadata (e.g., address, length, flag) from the device-controlled descriptors all suffer from the issue, which can result in memory corruption, and then lead to malicious code execution.

\begin{tcolorbox}
\begin{insight}
\label{insight1}
% In TEE middleware, host-facing interfaces are both inevitable and untrusted. Designs navigate trade-offs between performance, security, and TCB size: exposing more interfaces improves efficiency but widens the Iago attack surface, whereas stricter confinement strengthens isolation at the expense of flexibility.
In TEE middleware, host-facing interfaces are inevitable yet untrusted. Our study shows that different Tcons resolve this tension differently: some expose broader syscall and I/O surfaces for compatibility and performance, thereby enlarging the Iago attack surface; others enforce stricter confinement to reduce exposure, but at the cost of breaking legacy functionality and increasing development overhead.
\end{insight}
\end{tcolorbox}
\subsection{Insufficient I/O Security}
\label{sec:io_interface}

While memory inside TEEs is transparently encrypted by hardware, I/O operations are not. Thus, the responsibility of securing disk and file interactions falls onto the Tcon software. However, existing designs remain insufficient: even with encryption, leaks through syscalls and weaknesses in persistent storage still expose sensitive information. 

\subsubsection{Leaky I/O Metadata}

Process-based Tcons rely on the untrusted host kernel to handle I/O requests, inevitably exposing sensitive metadata to the OS. For example, during file operations, the kernel can infer details such as which file is being accessed, the specific offsets of read or write operations, and the access patterns. Even with encrypted filesystems, metadata leaks (such as file size, access timestamps, or offset patterns) can reveal sensitive insights, potentially compromising data confidentiality~\cite{chen2020metal}. For instance, if a database server runs on these Tcons, attackers may exploit offset information from read operations to learn access patterns or deduce which parts of a file are being accessed. Similar risks exist in network I/O, where unencrypted metadata such as packet length, block size, or timing patterns can leak application behavior and workload characteristics. Consequently, detailed information may still be inferred by an adversary~\cite{ahmad2018obliviate}.

Gramine employs a hybrid filesystem model~\cite{gramine_repo}, storing data on disk files with a specific encrypted format. However, operations on external files still expose access patterns to the kernel, undermining the confidentiality of sensitive workloads. Occlum, in contrast, implements measures to mitigate such threats. It employs an in-enclave filesystem to prevent file operations from being exposed to the untrusted OS. By caching upper-layer file I/O through its asynchronous filesystem (Async-SFS)~\cite{occlum_async_sfs}, Occlum both improves disk I/O performance and reduces metadata leakage, thereby enhancing security and anonymity.
Preventing such leakage is a cross-cutting issue for Tcons, spanning both disk and network I/O, and requires careful design beyond encryption alone.

\subsubsection{Broken Persistent Storage}

In the context of Tcons, writing data from memory to storage (e.g., hard disk) refers to transparent confidential persistent storage,
where we identified a number of potential vulnerabilities. 
Given that the host has full observability, attackers can replace any components outside the TEE.
Therefore in addition to ensuring availability and confidentiality, some primary attack vectors also must be addressed: (1) data corruption due to the absence of mounting integrity checks; and (2) replay attacks arising from a lack of freshness.

\vspace{3pt}\noindent\textbf{Lack of Availability or Confidentiality}.
In Mystikos, the ext2fs implementation relies on dm-verity, which is designed for read-only file systems and does not protect write operations. 
The lack of journaling in ext2fs means that data is not safeguarded against application crashes, with modifications lost when the application exits~\cite{mystikos_ext2fs}.
Hostfs, designed for data exchange between the LibOS and the host, operates without protection mechanisms~\cite{mystikos_hostfs}.
To strengthen Mystikos, enhancing the confidentiality and integrity mechanisms for all filesystem types is essential.

\vspace{3pt}\noindent\textbf{Lack of Mounting Integrity}.
In CoCo, a malicious host can exploit weaknesses by loading an encrypted container image onto the host and mounting it into the guest, where it is decrypted by the attestation agent~\cite{coco_issue_162}.
% https://github.com/confidential-containers/confidential-containers/issues/162
Once decrypted, the host gains visibility into sensitive data and can modify the container’s root filesystem, potentially altering workload behavior. This vulnerability stems from CoCo’s failure to validate mounted resources within the guest. To address this, the agent should verify volumes and include them in attestation evidence through security policies.

\begin{tcolorbox}[colback=white, colframe=black, boxrule=0.5pt, sharp corners]
\small
\textbf{Attack Vector 5: Snapshot Replay.}
Weak guarantees of atomicity and freshness let attackers tamper and replay with transient snapshots, enabling file rollback or corruption.
\end{tcolorbox}

\vspace{3pt}\noindent\textbf{Lack of Freshness and Atomicity}.
File operations in Occlum and Gramine lack atomicity and freshness, exposing critical vulnerabilities. Occlum's disk model allows adversaries to capture and replay transient snapshots, compromising application logic~\cite{occlum_snapshot}.
Similarly, Gramine’s Protected Files (PF) library~\cite{gramine_pfs} uses an in-enclave LRU cache, flushing intermediate files to disk when full. These truncated snapshots can be intercepted and manipulated, enabling attackers to tamper with critical files like ``redis.conf''. For example, missing fields such as \verb|requirepass| in truncated snapshots can let attackers bypass authentication.
Both systems fail to ensure atomic writes or validate snapshot freshness, enabling replay and truncation attacks. 
We reported them to the community, and a CVE was assigned to these vulnerabilities.

Addressing these vulnerabilities requires robust mechanisms to enforce atomicity and preserve snapshot integrity and freshness.
Intel released a patch~\cite{intel_patch} allowing larger cache sizes, reducing eviction frequency and thus transient snapshot generations.
Gramine also considers a pre-allocated free-list cache as mitigation~\cite{gramine_issue_1714}.
These minimal-engineering strategies are understandable but remain inadequate, as they still lack atomicity and freshness even with fewer transient snapshots.

\begin{tcolorbox}
\begin{insight}
\label{insight2}
% When TEE systems interact with the outside world, encryption is necessary but not sufficient. Beyond confidentiality, intermediate data must also be safeguarded for integrity, freshness, and atomicity. Without these guarantees, even encrypted I/O can be replayed, truncated, or reordered, undermining system correctness and security.
Encryption alone does not secure TEE I/O. Transient states such as snapshots, cache flushes, and intermediate files must be protected for integrity, freshness, and atomicity. Without these guarantees, adversaries can replay, truncate, or reorder encrypted data, leading to configuration bypasses, stale state, or corrupted application logic.
\end{insight}
\end{tcolorbox}

% https://sang.fan/assets/papers/rakis_eurosys25.pdf

\begin{table}[t]
\footnotesize
\begin{minipage}[t]{0.38\textwidth}
\centering
\caption{Vulnerabilities in Disk I/O}
\label{tab:storage_vulns}
\begin{tabular}{|c|c|}
\hline
Tcon & Issues \\ \hline
Gramine & Snapshot replay~\cite{gramine_encfiles} (CVE-2)\\ \hline
Occlum & Snapshot replay~\cite{code_release} (CVE-3)\\ \hline
Mystikos & Unencrypted hostfs~\cite{code_release} (Bug-7) \\ \hline
CoCo & Unverified mount~\cite{code_release,parma} (Bug-12) \\ \hline
\end{tabular}
\end{minipage}%
\hfill
\begin{minipage}[t]{0.58\textwidth}
\centering
\caption{Vulnerabilities in Orchestration Interfaces}
\label{tab:orchestration_vulns}
\begin{tabular}{|c|c|}
\hline
Component & Issues \\ \hline
Control plane & Wrong deployment from malicious controllers \\ \hline
CRIs & Access bypass, RCE, privilege escalation~\cite{cve_2024_21376} \\ \hline
Host commands & Unrestricted \verb|kubectl| exposing runtime state \\ \hline
OCI metadata & Malicious layers, tampered env/mounts~\cite{cve_2024_21403} \\ \hline
\end{tabular}
\end{minipage}
\end{table}
\subsection{Untrusted Orchestration Interfaces}
\label{sec:oi_interface}

As shown in Figure~\ref{fig:interface-design}, the orchestration flow from the control plane to the OCI-shim exposes multiple attack vectors that threaten the security of confidential and containerized environments. The unprotected container shim can introduce vulnerabilities by transmitting compromised container/VM images or guest agents during instantiation. Furthermore, the container runtime has the capability to alter or fabricate virtual hard disks, using these as layers for image construction, potentially embedding malicious code. Similarly, the shim can manipulate container filesystems by selecting arbitrary layers, injecting backdoors, or introducing environment variables and commands that exacerbate security risks. 
Additionally, remote communications remain vulnerable to interception or manipulation, facilitating man-in-the-middle attacks.

\subsubsection{Unprotected Control Plane and CRIs}

This subsection presents our empirical findings on security weaknesses introduced by Kubernetes-based orchestration in confidential container deployments. Through systematic boundary testing, we observe that while low-level TEE isolation remains intact, security guarantees are often undermined at higher layers by control-plane and Container Runtime Interface (CRI) components defined in the Kubernetes ecosystem. These externally exposed orchestration interfaces, though essential for large-scale management, become critical attack surfaces when their integrity and access control are insufficiently enforced.

\vspace{3pt}\noindent\textbf{Unprotected Control Plane}.
The Pod-centric security model in CoCo does not adequately safeguard high-level workload resources. If the Kubernetes control plane is compromised, the Pods may not function as intended by the users. For instance, when a new \verb|StatefulSet| is created, such as for a MySQL service, the controller is responsible for spawning a series of Pods with identical images but distinct numerical suffixes, ranging from `0' to `N'. The suffix determines whether the Pod is configured as a leader or a follower, which MySQL then uses to load its configuration. However, if an attacker were to maliciously initiate two Pods with the same suffix, such as `0', this would result in two leader Pods operating concurrently, thereby jeopardizing data integrity.
Such attacks occur when components in the control plane (e.g., controllers) are not protected, or their execution integrity is not verified.

\begin{tcolorbox}[colback=white, colframe=black, boxrule=0.5pt, sharp corners]
\small
\textbf{Attack Vector 6: Orchestration Interface Exploits
.}
Weak CRI protections enable access control bypass, RCE, or privilege escalation, while host-mediated APIs (e.g., settime) let adversaries disrupt application logic and compromise control-plane integrity.
\end{tcolorbox}

\vspace{3pt}\noindent\textbf{Unprotected CRIs}.
These vulnerabilities are meta-level issues that require careful consideration during the design and implementation of secure container orchestration systems.

Specific vulnerabilities highlight the insecurity of these CRIs. For instance, an improper access control flaw allows attackers to steal credentials and compromise resources beyond the intended security scope, including confidential guests and containers outside their designated network boundaries~\cite{cve_2024_29990}. Similarly, remote code execution and privilege escalation vulnerabilities let attackers gain unauthorized control to compromise the host and its dependent resources~\cite{cve_2024_21376}.

Moreover, untrusted timing mechanisms and host-controlled interfaces exemplify how the control plane and CRIs remain exposed. Adversaries can manipulate clocks, timers, or API calls to disrupt applications or exploit subtle inconsistencies. Our testing revealed that even trusted runtimes like Gramine, Mystikos, and Occlum can be affected by host-mediated timing errors~\cite{code_release}, underscoring the need to protect control plane components and their interfaces.
Furthermore, the \verb|settime| API exemplifies the risks of insecure design in CoCo~\cite{cross2024ccs}.

\vspace{3pt}\noindent\textbf{Unprotected Control Commands}.
Another issue is that control commands from the untrusted host are not isolated.
For example, the host can use `kubectl exec' to run arbitrary commands.
It can also solicit debugging information via `kubectl logs' from the running Tcons, which may include access to input/output operations, call stacks, or container-specific properties.

%OCI = config.json + rootfs
\subsubsection{Insecure OCI and Container Metadata}

Our analysis shows that insufficient protection of OCI artifacts and container metadata introduces a critical attack surface in Tcon deployments. Although application code is executed inside TEEs, key configuration files, image layers, and runtime parameters are often managed outside the trusted domain. As these metadata directly govern enclave initialization and execution semantics, their weak confidentiality and integrity guarantees enable adversaries to manipulate container behavior without violating hardware-enforced isolation.

\vspace{3pt}\noindent\textbf{Unencrypted Container Configuration}.
An adversary may engage in various activities that compromise the integrity and confidentiality of containerized environments, especially by manipulating the ``config.json'', 
thereby subverting the operational parameters.

Block devices, which store both immutable container image layers and the mutable scratch layer, are susceptible to unauthorized modifications. This can lead to the introduction of malicious code or the corruption of the container’s filesystem.
Attackers may manipulate container definitions in several ways: (1) reordering or injecting malicious layers into the overlay filesystem disrupts the integrity of the runtime environment; (2) modifying environment variables or altering user-specified commands allows adversaries to insert backdoors or introduce malicious behavior; (3) by modifying mount points, attackers can redirect container processes to compromised or unauthorized resources.
For instance, vulnerabilities~\cite{cve_2024_21403} reveal how improper files or directories within these layers can be exposed to unauthorized external access.

\vspace{3pt}\noindent\textbf{Lack of Configuration Content Validation}.
To ensure the security of applications running within enclaves, Tcons must protect not only the application but also its associated metadata. For arguments and environment variables with indeterminate content prior to enclave loading, security sanitization is essential.
An interesting case arises with Gramine-TDX, which moved its manifest file into the trusted domain. However, transferring configuration files into the VM remains a challenge. For instance, replacing insecure file-transfer methods like `kubectl cp' with alternatives like `scp' could mitigate these risks. 
Occlum incorporates arguments, environment variables, and configurations into its attestation process via initfs, ensuring their integrity, while also verifying and securely storing metadata for runtime-uploaded code.

\begin{tcolorbox}
\begin{insight}
\label{insight3}
% Tcon interfaces remain a key attack surface: unprotected control plane components, CRIs, host commands, and insecure OCI metadata can be exploited to manipulate container instantiation and execution.
Orchestration interfaces extend the Tcon attack surface: weak CRIs, unvalidated OCI metadata, and unprotected control-plane logic enable layer injection, command tampering, or scheduling subversion, undermining cluster-wide confidentiality and integrity.
\end{insight}
\end{tcolorbox}
\section{Lessons Learned and Mitigation Strategies}
\label{sec:tradeoffs}

\vspace{3pt}\noindent\textbf{Confidentiality, Atomicity, and Freshness of OS Interfaces}.
Tcon developers must minimize exposed OS interfaces, not only by reducing the number of I/O interfaces but also by sanitizing their parameters to ensure secure interaction with the underlying system. Developers should provide mechanisms for auditing or allow users to configure these interfaces to disable unnecessary calls, mitigating the risk of misuse and snooping.
For network I/O, enhanced solutions like Rkt-io~\cite{rkt_io} and Rakis~\cite{alharthi2025rakis} offer universal and transparent encryption across the I/O stack, ensuring both confidentiality and integrity of data entering and leaving the TEE. 
% This includes Layer 3 network packet encryption using Linux’s in-kernel Wireguard VPN~\cite{wireguard} for networking and full disk encryption with dm-crypt for storage. These mechanisms provide essential security guarantees for sensitive data, particularly in multi-tenant cases.

For disk I/O, some ongoing projects like MlsDisk~\cite{sworndisk} provide a virtual block device that enhances security while maintaining performance. We need a solution that can guarantee atomicity by ensuring that all writes are completed only upon a successful sync operation. 
CoCo ensure block-level integrity, using authentication tags for writable filesystems. However, these tags are vulnerable to replay attacks and do not ensure data freshness. The potential adoption of integrity trees, such as Merkle Trees, could address freshness concerns, but the cascading updates required along the root-leaf path impose significant latency and bandwidth overheads.
Future research should explore security-performance trade-offs in I/O to balance data freshness and operational atomicity.

\vspace{3pt}\noindent\textbf{Split APIs vs. Container Metadata Validation}.
Recent CoCo proposals aim to mitigate security risks but face critical limitations. One proposal introduces a policy engine to enforce tenant-defined configurations.
The container metadata validation~\cite{container_metadata_validation} provides attestation rooted in hardware-issued trust, enforcing user-specified execution policies. This mechanism extends the guest agent to ensure it only executes commands explicitly authorized by the tenant. The execution policy, defined and cryptographically measured by the user, is integrated into the attestation report~\cite{parma}.
However, this approach couples release policies tightly to specific execution policies, necessitating comprehensive revisions when container images require updates (e.g., for critical patches). While this rigid coupling enhances security, it may lack the flexibility some scenarios demand, where broader definitions of permissible actions are preferred. Future research could explore ways to maintain rigorous security guarantees while accommodating more dynamic policy definitions.

Another CoCo's proposal separates APIs into host-side (non-sensitive) and owner-side (sensitive)~\cite{splitapi,cross2024ccs}. More specifically, in-Pod command execution is classified as owner-side, while Pod creation/deletion is host-side. Despite this categorization, the approach fails to adequately demonstrate immunity to vulnerabilities across host-side APIs, leaving  attack surfaces unaddressed. 
% \weijie{sanitized but how?}

\vspace{3pt}\noindent\textbf{Confidential Kubernetes}.
The need for secure, multi-tenant Kubernetes solutions has driven the adoption of VM-based Tcon for private cluster protection and efficient resource management. 
% As TEE technologies mature, the focus can shift from TEE implementation to solving broader challenges, such as enforcing privacy compliance through data capsules~\cite{privguard}. 
These advancements position confidential orchestration frameworks as a new cornerstone.

CoCo provides minimal protection by isolating individual Pods within their TEEs, establishing a targeted security boundary while emphasizing lightweight modifications to the Kubernetes architecture. In contrast, confidential Kubernetes employs a monolithic protection model, encapsulating the entire Kubernetes stack within VM-based TEEs, as demonstrated by Constellation~\cite{constellation}. This model ensures the security of all cluster operations through trusted nodes and encrypted communications, leveraging advanced CNI plugins~\cite{cilium}.
These evolving approaches highlight the trade-offs between granularity and scalability in secure orchestration, paving the way for future innovations in confidential computing for cloud-native environments.

\section{Discussion}
\label{sec:discussion}

\vspace{3pt}\noindent\textbf{Limitation and Generalizability}.
This work primarily targets commodity x86-based TEE containers deployed in cloud environments.
As a result, the current implementation of TBouncer is not directly applicable to all TEE platforms or middleware systems.
Nevertheless, both our design principles and empirical findings have broader implications for the security analysis of TEE.
% \hongbo{why not say TEE here?}. 

At a high level, TBouncer is built upon a bidirectional synchronization and boundary-focused fuzzing model, which applies to TEE systems that expose well-defined interaction interfaces between trusted and untrusted components. 
CoCo, a CVM-based Tcon, exemplifies TBouncer's capabilities on supporting diverse architectures.
As TrustZone-based and SGX-based TEEs rely on Rich-OS/Trusted-OS and host/enclave boundaries, respectively, they are similar to the trust boundaries faced by TBouncer.
Prior studies have shown that this boundary is susceptible to Iago-style attacks~\cite{cerdeira2020sok}, which aligns with our findings in~\autoref{sec:os_interfaces}. 
Similar boundary-induced vulnerabilities have also been reported in TrustZone-based systems, where the untrusted Rich OS can manipulate inputs to the secure world or tamper with I/O responses returned to trusted components, thereby affecting trusted execution~\cite{lentz2018secloak}.
With appropriate enumeration of vendor-specific boundary interfaces, such as Secure Monitor Calls and GlobalPlatform APIs, TBouncer can be adapted to this setting to uncover such vulnerabilities systematically.
% Nevertheless, some systems (e.g., Intel SGX SDK) rely on interfaces implemented through specialized instructions (e.g., ECALL/OCALL) rather than standard OS abstractions. Supporting these systems in TBouncer requires substantial engineering effort.

% TEE middleware that requires source-level code changes, such as SGX SDK-based systems, relies on explicitly defined ECALL/OCALL interfaces implemented through specialized instructions rather than standard operating system abstractions. These systems follow a substantially different deployment from TEE containers, and are therefore not the primary focus of this work. 

\vspace{3pt}\noindent\textbf{Evaluation Metrics of TBouncer}.
Our results suggest that, for TEE middleware, assessing how thoroughly trust boundary interfaces are exercised is often more informative than relying solely on traditional fuzzing metrics such as code coverage.
In TEE environments, accurate coverage measurement is challenging due to limited observability, restricted instrumentation, and distinct runtime implementations. 
% More importantly, high code coverage does not necessarily imply thorough exploration of exposed interfaces that mediate interactions between trusted and untrusted components
We note that security properties unrelated to boundary enforcement, such as internal enclave logic, are orthogonal to our study and are not the primary targets of TBouncer.
Thus, as our primary focus is on the trust boundaries of Tcons rather than fuzzing the entire Tcon, code coverage veils the actual breadth of exploration on those interfaces.

% Overall, our results indicate that TBouncer achieves effective boundary exploration within a practical testing budget, making it suitable for routine security evaluation of TEE container deployments.
\section{Conclusion}
\label{sec:conclusion}

Our study shows that today’s Tcons provide insufficient protection and lack transparency regarding their security guarantees. We proposed and implemented a suite of benchmarks to systematically evaluate Tcons, uncovering new bugs (including 3 CVEs) and providing guidance for developers and users. We conclude with key takeaways and call for greater community attention to advancing techniques for securing current Tcons and informing future designs, hoping our findings serve as a foundation for the next generation of secure, transparent TEE containers.

\begin{acks}
% TODO: Replace with actual acknowledgments (funding, grants, etc.)
Weijie Liu, Shuo Huai, and Zheli Liu are supported by the National Natural Science Foundation of China under Grant No.62502237, No.U25B2028, and No.62432012.
\end{acks}

% \clearpage
\appendix
\section{Data availability}

% \section{Open Science}
The availability, functionality, and reproducibility of our artifact are ensured.
The source code is temporarily hosted on the website~\cite{code_release}.
% All code, CVEs, issues, and test inputs will be open-sourced after this paper's publication.
%
We reported our findings to the relevant communities and vendors (Intel, Gramine, Occlum, Mystikos, and CoCo), who have confirmed the existence of the risks we discovered. 
The examples provided in the paper have already been fixed, mitigated, discussed, or made public by the developers. 

% Our findings are listed on a dedicated website~\cite{code_release}, with reporter information omitted for anonymity and access restricted to reviewers with the provided link. Links to GitHub issues and CVE numbers will be made public after this paper's publication due to anonymity concerns. All testing was conducted responsibly on our local experimental environment, ensuring no harm to services or users of the affected software.

\clearpage

\bibliographystyle{ACM-Reference-Format}
\bibliography{ref}

\end{document}